\documentclass[12pt]{article}

\usepackage{amsmath, amsfonts, amsthm}
\usepackage{booktabs, multirow}
\usepackage{graphicx, pdfpages}
\usepackage{color, url}
\usepackage{natbib}
\usepackage{geometry, setspace}
\def\DE{{\mathrm{\scriptscriptstyle DE} }}
\def\IE{{\mathrm{\scriptscriptstyle IE} }}
\newcommand\independent{\protect\mathpalette{\protect\independenT}{\perp}}
\def\independenT#1#2{\mathrel{\rlap{$#1#2$}\mkern2mu{#1#2}}}

\DeclareMathOperator{\prev}{prev}
\DeclareMathOperator{\haz}{haz}
\allowdisplaybreaks[4]
\newtheorem{assumption}{Assumption}
\newtheorem{remark}{Remark}
\newtheorem{theorem}{Theorem}

\title{\bf Direct and Indirect Treatment Effects in the Presence of Semi-Competing Risks}

\author{Yuhao Deng$^1$, Yi Wang$^1$ and Xiao-Hua Zhou$^{1,2,*}$ \\
\small $^1$ Beijing International Center for Mathematical Research, Peking University \\
\small $^2$ Department of Biostatistics, School of Public Health, Peking Universiy}

\begin{document}

\maketitle

\begin{abstract}
Semi-competing risks refer to the phenomenon that the terminal event (such as death) can censor the non-terminal event (such as disease progression) but not vice versa. The treatment effect on the terminal event can be delivered either directly following the treatment or indirectly through the non-terminal event. We consider two strategies to decompose the total effect into a direct effect and an indirect effect under the framework of mediation analysis in completely randomized experiments by adjusting the prevalence and hazard of non-terminal events, respectively. They require slightly different assumptions on cross-world quantities to achieve identifiability. We establish asymptotic properties for the estimated counterfactual cumulative incidences and decomposed treatment effects. We illustrate the subtle difference between these two decompositions through simulation studies and two real-data applications. \par
{\bf Keywords:} Causal inference; Hazard; Markov; Mediation; Survival analysis.
\end{abstract}

\section{Introduction} \label{sec1}

Semi-competing risks involve non-terminal (intermediate) and terminal (primary) events. The non-terminal event can be censored by the terminal event, but not vice versa \citep{fine2001semi}. Compared with competing risks models, the analysis for semi-competing risks records the primary events even if intermediate events occur first. The dependence between the non-terminal event and terminal event was regarded as a fundamental problem in modeling the semi-competing rsks data \citep{xu2010statistical}. Typical methods to study semi-competing risks data include modeling the joint survivor function of the two events, especially using copula \citep{clayton1978model, fine2001semi, wang2003estimating, peng2007regression, hsieh2008regression, lakhal2008estimating} and modeling the data generating process using illness--death models \citep{wang2003nonparametric, xu2010statistical, lee2015bayesian, lee2016hierarchical}.

Despite their versatility, the aforementioned copula and illness--death model primarily focuses on estimating model parameters with limited causal implications. 
Causal analysis on semi-competing risks data aims to study the treatment effect on the primary event while appropriately adjusting the effect through the non-terminal event. Under the potential outcomes framework, there are generally two approaches to accommodating semi-competing risks. The first is principal stratification \citep{frangakis2002principal}, whose estimand is defined on a subpopulation classified by the joint values of two potential occurrences of non-terminal events under treated and under control \citep{nevo2022causal, xu2022bayesian, gao2022defining}. The second is mediation analysis \citep{baron1986moderator}, whose estimand is defined by contrasting potential distributions of primary events under the combination of hypothetical treatments associated with non-terminal event and primary event on the whole population \citep{vansteelandt2019mediation, huang2021causal}.

Especially, \citet{huang2021causal} proposed to decompose the total treatment effect on the terminal event into a natural direct effect (NDE) and a natural indirect effect (NIE). This work identified a promising direction to study causal effects with semi-competing risks data. Potential outcomes were defined on counting processes of data generating processes. NDE evaluates the treatment effect directly on the terminal event, and the NIE evaluates the treatment effect mediated by the non-terminal event. This work defined the causal estimand by contrasting counterfactual cumulative hazards. However, cumulative hazards are essentially integrals of conditional probabilities without intuitive interpretations. By some simple transformation, the difference in cumulative hazards is converted to the ratio of survival probabilities. Nevertheless, the ratio of probabilities is non-collapsible, in that one cannot easily summarize the treatment effect in the whole population by averaging those treatment effects in stratified subpopulations \citep{martinussen2013collapsibility}. A better way with intuitive interpretations could be to define the treatment effects as differences in counterfactual cumulative incidences \citep{buhler2023multistate}. Here, we define the cumulative incidence of the terminal event as the probability of experiencing a terminal event before a time point $t$, no matter whether there was a non-terminal event.

In this article, we first modify the causal estimand to the cumulative incidence scale, which is defined on the whole population and is collapsible. Then, we propose an alternative strategy to decompose the total effect. Instead of assuming the prevalence of non-terminal events is independent of cross-world treatment associated with the terminal event, we assume that the hazard of non-terminal events is independent of cross-world treatment associated with the terminal event. Although these two assumptions are both untestable, practitioners can choose a proper assumption by imagining the data generating mechanism and specifying the scientific question of interest. We provide asymptotic properties for the estimated counterfactual cumulative incidences and (direct and indirect) treatment effects. Simulation studies and real-data applications in Web Appendices illustrate the subtle difference between these two decompositions. Our proposed method can yield a conclusion consistent with clinical experience in some cases. 


\section{Framework and notations} \label{sec:framework}

Let $Z \in \{0, 1\}$ be a binary treatment, 1 for treated and 0 for control. Let $d\widetilde{N}_1(t; z_1)$ be the jump of the potential counting process of the non-terminal event during $[t, t+dt)$ when the treatment is set at $z_1$. Thus the potential counting process $\widetilde{N}_1(t; z_1) = \int_0^t d\widetilde{N}_1(s; z_1)$.
Let $d\widetilde{N}_2(t; z_2, n_1)$ be the jump of the potential counting process of the terminal event during $[t, t+dt)$ when the treatment is set at $z_2$ and the counting process of the non-terminal event at $t^{-}$ is set at $n_1$. Given a process $\tilde{n}_1(\cdot)$ of the non-terminal event, the potential counting process of the terminal event $\widetilde{N}_2(t; z_2, \widetilde{n}_1(\cdot)) = \int_0^t d\widetilde{N}_2(s; z_2, \widetilde{n}_1(s^-))$. Under this framework, we can firstly set the treatment at $z_1$ and get a potential process $\tilde{N}_1(\cdot;z_1)$ of the non-terminal event. Then we set the treatment at $z_2$ and supply the trajectory with $\tilde{N}_1(\cdot;z_1)$ to get a counterfactual process $\tilde{N}_2(\cdot;z_2,\tilde{N}_1(\cdot;z_1))$ of the terminal event.

To ensure that the intervention in $d\widetilde{N}_2(t; z_2, n_1)$ is well defined, we should assume that the hazard of the terminal event at time $t$ only depends on the status of the non-terminal event $n_1$ right before $t$.

\begin{assumption}[Markovness] \label{mar}
$P(d\widetilde{N}_2(t;z_2,n_1) = 1 \mid \widetilde{N}_2(t^-;z_2,\widetilde{n}_1(\cdot;z_1)) = 0, \widetilde{n}_1(t^-;z_1) = n_1, \widetilde{n}_1(s^-;z_1) = n_1^*) = P(d\widetilde{N}_2(t;z_2,n_1) = 1 \mid \widetilde{N}_2(t^-;z_2,\widetilde{n}_1(\cdot;z_1)) = 0, \widetilde{n}_1(t^-;z_1) = n_1)$ for $z_1, z_2, n_1, n_1^* \in \{0, 1\}$ and $0 < s < t < t^*$.
\end{assumption}

The potential time to the non-terminal event $T_1(z_1)$ given treatment assignment $z_1$ is the time that $\widetilde{N}_1(t; z_1)$ jumps. The potential time to the terminal event $T_2(z_1, z_2)$ given treatment assignment $z_2$ and intervened non-terminal event process $\widetilde{N}_1(\cdot;z_1)$ is the time that $\widetilde{N}_2(t; z_2, \tilde{N}_1(\cdot;z_1))$ jumps.
Rigorously, $\tilde{N}_1$ (or $T_1$) can be defined as a potential outcome of $z_1$ and $z_2$ since the non-terminal and termianl events are competing events. However, as we will see in the next section, there is no need to step into the counterfactual world for $\tilde{N}_1$ (or $T_1$) for identification of treatment effects.
Let $\tilde{N}_1(t)$ and $\tilde{N}_2(t)$ be the observable counting processes of the non-terminal and terminal event in the absence of censoring. Let $C$ be the censoring time and $t^*$ be the end of study. The following assumptions are standard in causal inference.

\begin{assumption}[Causal consistency] \label{con}
$d\widetilde{N}_1(t) = d\widetilde{N}_1(t; Z)$ and $d\widetilde{N}_2(t) = d\widetilde{N}_2(t; Z, \widetilde{N}_1(\cdot;Z))$.
\end{assumption}

\begin{assumption}[Ignorability] \label{ign}
$\{d\widetilde{N}_1(t^-; z_1), d\widetilde{N}_2(t; z_2,n_1), 0 < t < t^*\} \independent Z$ for $z_1, z_2, n_1 \in \{0, 1\}$.
\end{assumption}

\begin{assumption}[Random censoring] \label{ran}
$\{\widetilde{N}_1(t; z), \widetilde{N}_2(t; z, \widetilde{N}_1(\cdot;z)): 0 < t < t^*\} \independent C \mid Z=z$ for $z, n_1 \in \{0, 1\}$.
\end{assumption}

\begin{assumption}[Positivity] \label{pos}
(i) If $P(\tilde{N}_1(t;z_1)=n_1, \tilde{N}_2(t;z_2,\tilde{N}_1(\cdot;z_1))=0) > 0$, then $P(\tilde{N}_1(t)=n_1, \tilde{N}_2(t)=0, Z=z) > 0$, for $z,z_1,z_2,n_1\in\{0,1\}$.
(ii) $P(C > t^* \mid Z) > 0$.
\end{assumption}

Ignorability (Assumption \ref{ign}) states that the treatment assignment is independent of all potential values. Random censoring (Assumption \ref{ran}) states that the censoring time is independent of potential failure times, which is equivalent to $\{T_1(z), T_2(z,z)\} \independent C \mid Z=z$. Here we treat the censoring time as a non-potential value, although we allow $C$ to be dependent on the treatment $Z$. It is straightforward to write $C$ as a potential outcome of $Z$ with ignorability, but it does not make any difference in identification or estimation for the target causal estimand. Positivity (Assumption \ref{pos}) ensures that there are data to evaluate the distribution as long as the counterfactual trajectory of events is possible, and there are units still at risk at the end of the study.

Like classical mediation analysis, a sequential ignorability should be employed \citep{imai2010identification}. Given the history of the treatment and the status of events, the jump of the terminal event should have no cross-world reliance on the potential process of the non-terminal event. Whether the terminal event happens at time $t$ is independent of the potential non-terminal event process after $t$. This assumption generally means that there is no confounding for the status of the non-terminal event and instantaneous jump of the terminal event.

\begin{assumption}[Sequential ignorability, part 1] \label{seq}
$P(d\widetilde{N}_2(t; z_2, n_1) = 1 \mid Z = z_2, \widetilde{N}_1(t^-; z_1) = n_1, \widetilde{N}_2(t^-; z_2, \widetilde{N}_1(\cdot;z_1)) = 0) = P(d\widetilde{N}_2(t; z_2, n_1) = 1 \mid Z = z_2, \widetilde{N}_1(t^-) = n_1, \widetilde{N}_2(t^-) = 0)$ for $z_1, z_2, n_1 \in \{0, 1\}$ and $0 < t < t^*$.
\end{assumption}

We introduce some notations for observed processes. Let $T_1$ be the time to non-terminal event, and $T_2$ the time to terminal event. Under causal consistency, $T_1 = T_1(Z)$ and $T_2 = T_2(Z,Z)$. The censoring indicators for the non-terminal event and terminal event are $\delta_1 = I\{T_1 \le C\}$ and $\delta_2 = I\{T_2 \le C\}$, respectively. The observed counting process and at-risk process for the non-terminal event are $N_{*}(t) = I\{T_1 \le t, T_2 \ge T_1, \delta_1 = 1\}$ and $Y_{*}(t) = I\{T_1 \ge t, T_2 \ge t, C \ge t\}$, respectively. The observed counting process and at-risk process for the terminal event without prior non-terminal event are $N_0(t) = I\{T_2 \le t, T_1 \ge T_2, (1-\delta_1) \delta_2 = 1\}$ and $Y_0(t) = I\{T_1 \ge t, T_2 \ge t, C \ge t\}$, respectively. The observed counting process and at-risk process for the terminal event with prior non-terminal event are $N_1(t) = I\{T_1 \le t, T_2 \le t, \delta_1 \delta_2 = 1\}$ and $Y_1(t) = I\{T_1 < t, T_2 \ge t, C \ge t\}$, respectively. The at-risk processes $Y_{*}(t) = Y_{0}(t)$ because the non-terminal event and direct terminal event (without prior non-terminal event) are a pair of competing events, sharing the same at-risk set.

\begin{remark}
When there are ties for the non-terminal and terminal events, i.e., $T_1(z_1) = T_2(z_1,z_2)$, we assume the non-terminal event happens just before the terminal event. This consideration is meaningful in clinical studies with disease progression. For example, to investigate the effect of stem cell transplantation, the non-terminal event is relapse, and the terminal event is death. Death without relapse is called non-relapse mortality or treatment-related mortality, whereas death with relapse is called relapse-related mortality. If death and relapse happen at the same time, such a death event should be classified as relapse-related mortality.
\end{remark}

Suppose the sample size is $m$ in a randomized controlled trial. We observe $m$ independent and identically distributed copies of $\{Z, N_{n}(t), Y_{n}(t), \delta_1, \delta_2: 0 \le t \le t^*, n = *, 0, 1\}$. Equivalently, the observed data for each individual include $\{Z, T_1\wedge C, \delta_1, T_2\wedge C, \delta_2\}$. We use the subscript $i$ to denote the observation of the $i$th individual. Let $\overline{N}_{n}(t; z) = \sum_{i=1}^{m} I\{Z_i=z\}N_{n,i}(t)$ and $\overline{Y}_{n}(t; z) = \sum_{i=1}^{m} I\{Z_i=z\}Y_{n,i}(t)$, $n = *, 0, 1$.

We adopt the counterfactual cumulative incidence of the terminal event $$F(t; z_1, z_2) = P(\widetilde{N}_2(t; z_2, \widetilde{N}_1(\cdot;z_1)) = 1)$$ as the quantity of interest, because cumulative incidences have intuitive causal interpretations and are collapsible.
The total treatment effect $\Delta(t) = F(t; 1, 1) - F(t; 0, 0)$ measures the combination of a natural direct effect (NDE) on the terminal event $\Delta_{\DE}(t) = F(t; 0, 1) - F(t; 0, 0)$ and a natural indirect effect (NIE) on the terminal event through the non-terminal event $\Delta_{\IE}(t) = F(t; 1, 1) - F(t; 0, 1)$. In NDE, the causal pathway from treatment to non-terminal event is controlled. In NIE, the causal pathway into terminal event is controlled. Heuristically, NDE explains the treatment effect on the cumulative incidence of the terminal event by affecting the ``risk'' of the terminal event, and NIE explains that by affacting the ``risk'' of the non-terminal event.

To identify the natural direct effect and natural indirect effect, it is equivalent to identify the counterfactual hazard of the terminal event
\begin{align*}
\frac{d}{dt}\Lambda(t; z_1, z_2) &= - \frac{d}{dt} \log \{1 - F(t; z_1, z_2)\} \\
&= \frac{1}{dt} P(d\widetilde{N}_2(t; z_2, \widetilde{N}_1(t^-; z_1)) = 1 \mid \widetilde{N}_2(t^-; z_2, \widetilde{N}_1(\cdot;z_1)) = 0)
\end{align*}
because there is a one-to-one relationship between the hazard and cumulative incidence.

\section{Natural direct and indirect effects} \label{sec:deie}

\subsection{Identification of the hazards of terminal events}

There are two hazards for the terminal event, reflecting the instantaneous risk of the terminal event with a non-terminal event and without a non-terminal event, respectively. Define the counterfactual hazards of the terminal event by intervening the treatment at $z_2$ and the counting process of the non-terminal event at $\widetilde{N}_1(\cdot;z_1)$ as
\[
\frac{d}{dt}\Lambda_{n_1}(t; z_1, z_2) = \frac{1}{dt} P(d\widetilde{N}_2(t; z_2, n_1) = 1 \mid \widetilde{N}_1(t^-; z_1) = n_1, \widetilde{N}_2(t^-; z_2, \widetilde{N}_1(\cdot;z_1)) = 0),
\]
$n_1 = 0, 1$.
Under Assumptions \ref{con}--\ref{ran}, $\Lambda_{n_1}(t; z_1, z_2) \equiv \Lambda_{n_1}(t; z_2)$; see Web Appendix A.
It can be estimated by Nelson-Aalen estimator,
\[
d\widehat\Lambda_{n_1}(t; z_2) = \frac{I\{\overline{Y}_{n_1}(t; z_2) > 0\}}{\overline{Y}_{n_1}(t; z_2)}d\overline{N}_{n_1}(t; z_2), \quad n_1 = 0, 1, \ z_2 = 0, 1.
\]


Although we have shown that the hazards of the terminal event are identifiable and estimable, we still need additional assumptions to identify quantities in the world involving the treatment associated with the non-terminal event. Different assumptions may reflect different strategies to interpret treatment effects.

\subsection{Decomposition 1: Controlling the prevalence of non-terminal events}

To decompose the total effect into a direct and an indirect effect, \citet{huang2021causal} introduced the following assumption on the prevalence of non-terminal events.

\begin{assumption}[Sequential ignorability, part 2: controlling the prevalence] \label{pre}
$P(\widetilde{N}_1(t; z_1) = 1 \mid Z = z_1, \widetilde{N}_2(t; z_2, \widetilde{N}_1(\cdot;z_1)) = 0) = P(\widetilde{N}_1(t; z_1) = 1 \mid Z = z_1, \widetilde{N}_2(t) = 0)$, for $z_1, z_2 \in \{0, 1\}$ and $0 < t < t^*$.
\end{assumption}

Define the prevalence of the non-terminal event by intervening the treatment at $z_2$ and counting process of the non-terminal event at $\widetilde{N}_1(\cdot;z_1)$ as
\[
w_{n_1}(t; z_1, z_2) = P(\widetilde{N}_1(t; z_1) = n_1 \mid \widetilde{N}_2(t; z_2, \widetilde{N}_1(\cdot;z_1)) = 0).
\]
Assumption \ref{pre} states that the prevalence of non-terminal events does not rely on the potential process of the terminal event as long as the terminal event has not occurred. Under this assumption, the prevalence of non-terminal events $w_{n_1}(t; z_1, z_2) \equiv w_{n_1}(t; z_1)$; see Web Appendix A. This prevalence can be estimated by
\[
\widehat{w}_{n_1}(t; z_1) = \frac{I\{\overline{Y}_0(t; z_1) + \overline{Y}_1(t; z_1) > 0\}}{\overline{Y}_0(t; z_1) + \overline{Y}_1(t; z_1)} \overline{Y}_{n_1}(t; z_1), \quad n_1 = 0, 1, \  z_1 = 0, 1.
\]

We have the following result to identify the counterfactual hazard of the terminal event.


\begin{theorem}
Under Assumptions \ref{mar}--\ref{pos} and \ref{pre},
\[
F(t;z_1,z_2) = 1 - \exp\left\{-\int_0^t \sum_{n_1 \in \{0,1\}} w_{n_1}(s^-;z_1) d\Lambda_{n_1}(s;z_2)\right\}
\]
is identifiable. Estimators of $F(t; z_1, z_2)$, $\Delta_{\DE}(t)$, $\Delta_{\IE}(t)$ and their asymptotic properties are provided in Web Appendix A.
\end{theorem}

\subsection{Decomposition 2: Controlling the hazard of non-terminal events}

In some scenarios, interpreting the natural direct effect as ``controlling the prevalence of non-terminal events'' may not be reasonable. For example, if a novel therapy almost removes terminal events, we may expect that the direct effect should be large but the indirect effect is absent. However, since the prevalence of non-terminal events increases greatly by removing terminal events, the indirect effect is also present. Therefore, we replace Assumption \ref{pre} which controls prevalence with the following assumption which controls hazard.

\begin{assumption}[Sequential ignorability, part 2: controlling the hazard] \label{haz}
$P(d\widetilde{N}_1(t; z_1) = 1 \mid Z = z_1, \widetilde{N}_1(t^-; z_1) = 0, \widetilde{N}_2(t^-; z_2, \widetilde{N}_1(\cdot;z_1)) = 0) = P(d\widetilde{N}_1(t; z_1) = 1 \mid Z = z_1, \widetilde{N}_1(t^-) = 0, \widetilde{N}_2(t^-) = 0)$, for $z_1, z_2 \in \{0, 1\}$ and $0 < t < t^*$.
\end{assumption}

Define the hazard of the non-terminal event by intervening the treatment at $z_2$ and counting process of thenon-terminal event at $\widetilde{N}_1(\cdot;z_1)$ as
\[
\frac{d}{dt}\Lambda_{*}(t; z_1, z_2) = \frac{1}{dt}P(d\widetilde{N}_1(t; z_1) = 1 \mid \widetilde{N}_1(t; z_1) = 0, \widetilde{N}_2(t; z_2, \widetilde{N}_1(\cdot;z_1)) = 0).
\]
Assumption \ref{haz} states that the hazard of the non-terminal event does not rely on the potential process of the terminal event as long as the terminal event has not occurred. Under Assumptions \ref{con}--\ref{ran} and \ref{haz}, $\Lambda_{*}(t; z_1, z_2) \equiv \Lambda_{*}(t; z_1)$; see Web Appendix B. It can be estimated by Nelson-Aalen estimator,
\[
d\widehat\Lambda_{*}(t; z_1) = \frac{I\{\overline{Y}_{*}(t; z_1) > 0\}}{\overline{Y}_{*}(t; z_1)} d\overline{N}_{*}(t; z_1), \quad z_1 = 0, 1.
\]


\begin{theorem}
Under Assumptions \ref{mar}--\ref{pos} and \ref{haz},
\begin{align*}
F(t;z_1,z_2) &= \int_0^t \exp\{-\Lambda_{*}(s^-; z_1) - \Lambda_0(s^-; z_2)\} \cdot d\Lambda_0(s; z_2) \\
&\quad + \int_0^{t^-} \exp\{-\Lambda_{*}(s^-; z_1) - \Lambda_0(s^-; z_2)\} [1 - \exp\{-\Lambda_1(t; z_2) + \Lambda_1(s; z_2)\}] d\Lambda_{*}(s; z_1)
\end{align*}
is identifiable. Estimators of $F(t; z_1, z_2)$, $\Delta_{\DE}(t)$, $\Delta_{\IE}(t)$ and their asymptotic properties are provided in Web Appendix B.
\end{theorem}

\section{Discussion} \label{sec:discussion}

This article considers two decompositions for the total effect in the presence of semi-competing risks. Heuristically, the direct effect should reflect the treatment effect on the terminal event by appropriately controlling the ``risk'' of the non-terminal event, and the indirect effect should reflect the treatment effect on the terminal event by changing the ``risk'' of the non-terminal event while controlling the hazards of the terminal event. Sequential ignorability (part 2) puts forward a question: what kind of ``risk'' should be controlled? When shifting the treatment associated with the terminal event, does the prevalence or hazard of the non-terminal event remain unchanged? In Web Appendix C, we conduct simulation studies to compare these two decompositions in several settings. In Web Appendix D, we conduct real-data applications on a hepatitis B dataset and a leukemia dataset. These two decompositions can lead to either similar or different results, according to specific data scenarios. In some scenarios, the conclusion obtained by our proposed decomposition is more in line with practical experience. 

Decomposition 2 can be better understood by utilizing a multi-state model. Terminal events consist of two parts: direct outcome events without a prior non-terminal event, and indirect outcome events with a prior non-terminal event. Hazard functions correspond to transition rates between states. The three estimated hazards (of direct outcome event, non-terminal event and indirect outcome event) are independent, so they can be tested separately based on logrank statistics. Practitioners can easily know on which pathway the treatment effect is present. If the hazard of the non-terminal event remains identical under treatment and under control, then the natural indirect effect is absent. If the two hazards of the terminal event remain identical under treatment and under control, then the natural direct effect is absent. By weighted logrank tests and intersection-union test, \citet{huang2022hypothesis} proposed methods to test the natural indirect effect under Decomposition 1.

Decomposition 2 is related to the separable effects approach in causal inference \citep{stensrud2021generalized, stensrud2022separable, breum2024estimation}. The separable effects approach simplifies the notations by assuming that the treatment has dismissible components, each of which only affects one pathway. Treatment effects due to dismissible components on terminal and non-terminal events are assumed to be isolated. In this way, we can decompose the total treatment effect into three separable effects. Assumptions under the separable effects framework are literally different but essentially similar to those under the mediation analysis framework. The separable effect for the pathway from non-terminal event to terminal event serves as an interaction effect beyond the direct and indirect effects.

\section*{Acknowledgments}

Funding information: National Key Research and Development Program of China, Grant No.~2021YFF0901400; National Natural Science Foundation of China, Grant No.~12026606, 12226005. This work is also partly supported by Novo Nordisk A/S.

\bibliographystyle{apalike}
\bibliography{ref}

\newpage

\appendix

\renewcommand\thesection{\Alph{section}}
\renewcommand {\theequation} {S\arabic{equation}}
\renewcommand {\thetable} {S\arabic{table}}
\renewcommand {\thefigure} {S\arabic{figure}}

\begin{center}
\section*{Appendix}
\end{center}

\section{Proof of Theorem 1}

\subsection{Identification}

Under Assumptions 2, 3, 4 and 6,
\begin{align*}
\frac{d}{dt}\Lambda_{n_1}(t; z_1, z_2) &:= \frac{1}{dt} P(d\widetilde{N}_2(t; z_2, n_1) = 1 \mid \widetilde{N}_1(t^-; z_1) = n_1, \widetilde{N}_2(t^-; z_2, \widetilde{N}_1(\cdot;z_1)) = 0) \\
&= \frac{1}{dt} P(d\widetilde{N}_2(t; z_2, n_1) = 1 \mid Z = z_2, \widetilde{N}_1(t^-; z_1) = n_1, \widetilde{N}_2(t^-; z_2, \widetilde{N}_1(\cdot;z_1)) = 0) \\
&= \frac{1}{dt} P(d\widetilde{N}_2(t; z_2, n_1) = 1 \mid Z = z_2, \widetilde{N}_1(t^-; z_2) = n_1, \widetilde{N}_2(t^-; z_2, \widetilde{N}_1(\cdot;z_2)) = 0) \\
&= \frac{1}{dt} P(d\widetilde{N}_2(t; z_2, n_1) = 1 \mid Z = z_2, \widetilde{N}_1(t^-; z_2) = n_1, \widetilde{N}_2(t^-; z_2, \widetilde{N}_1(\cdot;z_2)) = 0, C \ge t) \\
&= \frac{1}{dt} P(d\widetilde{N}_2(t) = 1 \mid Z = z_2, \widetilde{N}_1(t^-) = n_1, \widetilde{N}_2(t^-) = 0, C \ge t) \\
&= \frac{1}{dt} P(dN_2(t) = 1 \mid Z = z_2, N_1(t^-) = n_1, Y_2(t) = 0).
\end{align*}
This implies that $d\Lambda_{n_1}(t; z_1, z_2) = d\Lambda_{n_1}(t; z_2, z_2) := d\Lambda_{n_1}(t; z_2)$. Assumption 5 guarantees the conditional probability is estimable.

Under Assumption 7,
\begin{align*}
w_{n_1}(t; z_1, z_2) &:= P(\widetilde{N}_1(t; z_1) = n_1 \mid \widetilde{N}_2(t; z_2, \widetilde{N}_1(\cdot;z_1)) = 0)) \\
&= P(\widetilde{N}_1(t; z_1) = n_1 \mid Z = z_1, \widetilde{N}_2(t; z_2, \widetilde{N}_1(\cdot;z_1)) = 0)) \\
&= P(\widetilde{N}_1(t; z_1) = n_1 \mid Z = z_1, \widetilde{N}_2(t; z_1, \widetilde{N}_1(\cdot;z_1)) = 0)) \\
&= w_{n_1}(t; z_1, z_1).
\end{align*}
This implies that $w_{n_1}(t; z_1, z_1) = w_{n_1}(t; z_1, z_2) := w_{n_1}(t; z_1)$.

Therefore, the counterfactual hazard of the terminal event
\begin{align*}
d\Lambda(t;z_1,z_2) &:= P(d\widetilde{N}_2(t; z_2, \widetilde{N}_1(\cdot;z_1)) = 1 \mid \widetilde{N}_2(t^-; z_2, \widetilde{N}_1(\cdot;z_1)) = 0) \\
&= \sum_{n_1\in\{0,1\}} P(d\widetilde{N}_2(t; z_2, \widetilde{N}_1(\cdot;z_1)) = 1 \mid \widetilde{N}_1(t^-; z_1) = n_1, \widetilde{N}_2(t^-; z_2, \widetilde{N}_1(\cdot;z_1)) = 0) \\
&\qquad\qquad \cdot P(\widetilde{N}_1(t^-; z_1)= n_1 \mid \widetilde{N}_2(t^-; z_2, \widetilde{N}_1(\cdot;z_1)) = 0) \\
&= \sum_{n_1\in\{0,1\}} d\Lambda_{n_1}(t; z_1, z_2) \cdot w_{n_1}(t^-; z_1, z_2).
\end{align*}
And thus the counterfactual cumulative incidence $F(t;z_1,z_2)$ is identifiable, denoted by
\begin{align*}
F_{\prev}(t;z_1,z_2) &:= 1 - \exp\{-\Lambda(t;z_1,z_2)\} \\
&= 1 - \exp\left\{-\int_0^t \sum_{n_1 \in \{0,1\}} d\Lambda_{n_1}(s;z_1,z_2) \cdot w_{n_1}(s^-;z_1,z_2)\right\} \\
&= 1 - \exp\left\{-\int_0^t \sum_{n_1 \in \{0,1\}} d\Lambda_{n_1}(s;z_2) \cdot w_{n_1}(s^-;z_1)\right\}.
\end{align*}

A consistent estimator for the counterfactual cumulative incidence is given by
\[
\widehat{F}_{\prev}(t;z_1,z_2) = 1 - \exp\left\{-\int_0^t \sum_{n_1 \in \{0,1\}} d\widehat\Lambda_{n_1}(s;z_2) \cdot \widehat{w}_{n_1}(s^-;z_1)\right\},
\]
where
\begin{align*}
d\widehat\Lambda_{n_1}(t; z_2) &= \frac{I\{\overline{Y}_{n_1}(t; z_2) > 0\}}{\overline{Y}_{n_1}(t; z_2)}d\overline{N}_{n_1}(t; z_2), \quad n_1 = 0, 1, \ z_2 = 0, 1, \\
\widehat{w}_{n_1}(t; z_1) &= \frac{I\{\overline{Y}_0(t; z_1) + \overline{Y}_1(t; z_1) > 0\}}{\overline{Y}_0(t; z_1) + \overline{Y}_1(t; z_1)} \overline{Y}_{n_1}(t; z_1), \quad n_1 = 0, 1, \  z_1 = 0, 1,
\end{align*}
with $\overline{N}_{n_1}(t; z_2)$ and $\overline{Y}_{n_1}(t; z_1)$ defined in Section 2.

\subsection{Estimated counterfactual cumulative incidence}

We first derive the asymptotic property for the estimator of the counterfactual cumulative hazard $\Lambda(t;z_1,z_2)$, denoted by $\widehat\Lambda(t;z_1,z_2)$. Denote
\[
\overline{M}_{n_1}(t;z_2) = \int_0^t d\overline{N}_{n_1}(s;z_2) - \overline{Y}_{n_1}(s;z_2) d\Lambda_{n_1}(s;z_2).
\]
Under Assumption 1 (Markovness), $\overline{M}_{n_1}(t;z_2)$ is a martingale with respect to $\overline{N}_{n_1}(s;z_2)$, where $n_1 = 0, 1$. Hereafter we simplify $\overline{Y}_{n_1}^{-1}(s;z_2)I\{\overline{Y}_{n_1}(s;z_2)>0\}$ as $\overline{Y}_{n_1}^{-1}(s;z_2)$ because positivity ensures that $\overline{Y}_{n_1}(s;z_2)>0$ almost surely.
\begin{align*}
&\quad ~ \widehat\Lambda(t;z_1,z_2) - \Lambda(t;z_1,z_2) \\
&= \sum_{n_1\in\{0,1\}} \int_0^t \widehat{w}_{n_1}(s^-;z_1) d\widehat\Lambda_{n_1}(s;z_2) - \sum_{n_1\in\{0,1\}} \int_0^t w_{n_1}(s^-;z_1) d\Lambda_{n_1}(s;z_2) \\
&= \sum_{n_1\in\{0,1\}} \int_0^t w_{n_1}(s^-;z_1) d\widehat\Lambda_{n_1}(s;z_2) + \sum_{n_1\in\{0,1\}} \int_0^t \{\widehat{w}_{n_1}(s^-;z_1) - w_{n_1}(s^-;z_1)\} d\widehat\Lambda_{n_1}(s;z_2) \\
&\qquad - \sum_{n_1\in\{0,1\}} \int_0^t w_{n_1}(s^-;z_1) d\Lambda_{n_1}(s;z_2) \\
&= \sum_{n_1\in\{0,1\}} \int_0^t w_{n_1}(s^-;z_1) \overline{Y}_{n_1}^{-1}(s;z_2) d\overline{M}_{n_1}(s;z_2) \\
&\qquad + \sum_{n_1\in\{0,1\}} \int_0^t \{\widehat{w}_{n_1}(s^-;z_1) - w_{n_1}(s^-;z_1)\} \{d\Lambda_{n_1}(s;z_2) + \overline{Y}_{n_1}^{-1}(s;z_2)d\overline{M}_{n_1}(s;z_2)\} \\
&= \sum_{n_1\in\{0,1\}} \int_0^t w_{n_1}(s^-;z_1)\overline{Y}_{n_1}^{-1}(s;z_2) d\overline{M}_{n_1}(s;z_2) \\
&\qquad + \sum_{n_1\in\{0,1\}} \int_0^t \{\widehat{w}_{n_1}(s^-;z_1) - w_{n_1}(s^-;z_1)\} d\Lambda_{n_1}(s;z_2) + o_p(m^{-1/2}).
\end{align*}
The first term is a sum of two martingales because $w_{n_1}(s^-;z_1)Y_{n_1}^{-1}(s;z_2)$ is predictable. The martingale processes for $n_1 = 0$ and $n_1 = 1$ are independent because they jump at different times with probability 1. We have that $m^{-1}\overline{Y}_{n_1}(s;z_2)$ converges to $P(T_2 \wedge C \ge s, \widetilde{N}_1(s^-)=n_1, Z=z_2)$ by law of large numbers and $m^{-1/2}\overline{M}_{n_1}(s;z_2)$ converges to a Gaussian process (also a martingale) with mean zero and variance $P(T_2 \wedge C \ge s, \widetilde{N}_1(s^-)=n_1, Z=z_2) \Lambda_{n_1}(s;z_2)$ by martingale central limit theroem. For the second term, ${m}^{1/2}\{\widehat{w}_{n_1}(s^-;z_1) - w_{n_1}(s^-;z_1)\}$ converges to a Gaussian process $X_{n_1}(s^-;z_1)$ with mean zero and variance provided in \citet{huang2021causal}. In addition, these two terms are independent because $\overline{M}_{n_1}(s;z_2)$ and $\widehat{w}_{n_1'}(s^-;z_1)$ jump at different times (the former jumps at $T_2\wedge C$ but the latter does not). Applying the continuous mapping theorem,
\begin{align*}
&\quad m^{1/2}\{\widehat\Lambda_{\prev}(t;z_1,z_2) - \Lambda_{\prev}(t;z_1,z_2)\} \\
& \xrightarrow{d} \sum_{n_1\in\{0,1\}} \int_0^t \frac{w_{n_1}(s^-;z_1)}{P(T_2 \wedge C \ge s, \widetilde{N}_1(s^-)=n_1, Z=z_2)} d\overline{M}_{n_1}(s;z_2) \\
&\qquad + \sum_{n_1\in\{0,1\}} \int_0^t X_{n_1}(s^-;z_1) d\Lambda_{n_1}(s;z_2) \\
&= \sum_{n_1\in\{0,1\}} \int_0^t \frac{w_{n_1}(s^-;z_1)}{P(T_2 \wedge C \ge s, \widetilde{N}_1(s^-)=n_1, Z=z_2)} d\overline{M}_{n_1}(s;z_2) \\
&\qquad + \int_0^t X_{1}(s^-;z_1) \{d\Lambda_{1}(s;z_2)-d\Lambda_{0}(s;z_2)\}
\end{align*}
by noting that $\{\widehat{w}_{1}(s^-;z_1) - w_{1}(s^-;z_1)\} = - \{\widehat{w}_{0}(s^-;z_1) - w_{0}(s^-;z_1)\}$. Finally,
\[
m^{1/2} \left\{\widehat\Lambda(t;z_1,z_2) - \widehat\Lambda(t;z_1,z_2)\right\} \xrightarrow{d} N\{0, \sigma^2_{\prev}(t;z_1,z_2)\},
\]
where
\begin{align*}
\widehat\sigma^2_{\prev}(t;z_1,z_2)
&= \sum_{n_1\in\{0,1\}} \int_0^t \frac{{w}_{n_1}(s^-;z_1)^2}{P(T_2 \wedge C \ge s, \widetilde{N}_1(s^-)=n_1, Z=z_2)}  d\Lambda_{n_1}(s;z_2) \\
&\qquad + E\left[\int_0^t X_{1}(s^-;z_1) \{d\Lambda_{1}(s;z_2) - d\Lambda_{0}(s;z_2)\}\right]^2.
\end{align*}
Then using delta method,
\begin{align*}
m^{1/2} \{\widehat{F}_{\prev}(t;z_1,z_2) - F(t;z_1,z_2)\} \xrightarrow{d} N\left\{0, \ \exp\{-2\Lambda(t;z_1,z_2)\} \sigma_{\prev}^2(t;z_1,z_2)\right\}.
\end{align*}

The prevalence of non-terminal events is held unchanged when calculating the natural direct effect. Under this decomposition, NDE and NIE are estimated by $\widehat\Delta_{\DE}(t) = \widehat{F}_{\prev}(t;0,1) - \widehat{F}_{\prev}(t;0,0)$ and $\widehat\Delta_{\IE}(t) = \widehat{F}_{\prev}(t;1,1) - \widehat{F}_{\prev}(t;0,1)$, respectively. NDE measures the treatment effect on the cumulative incidence of terminal events via changing the hazards of terminal events while controlling the prevalence of non-terminal events. NIE measures the treatment effect on the cumulative incidence of terminal events via changing the prevalence of terminal events while controlling the hazards of terminal events.

\subsection{Estimated natural direct effect}

The natural direct effect (NDE) is estimated by
\begin{align*}
\widehat\Delta_{\DE}(t) &= \widehat{F}_{\prev}(t;0,1) - \widehat{F}_{\prev}(t;0,0) \\
&= \exp\left\{-\widehat\Lambda(t;0,0)\right\} - \exp\left\{-\widehat\Lambda(t;0,1)\right\}.
\end{align*}
To show its asymptotic law,
\begin{align*}
\widehat\Delta_{\DE} - \Delta_{\DE} &= \exp\left\{-\widehat\Lambda(t;0,0)\right\} - \exp\left\{-\Lambda(t;0,0)\right\} - \exp\left\{-\widehat\Lambda(t;0,1)\right\} + \exp\left\{-\Lambda(t;0,1)\right\} \\
&= \exp\left\{-\Lambda(t;0,0)\right\} \left[\widehat\Lambda(t;0,0)-\Lambda(t;0,0)\right] \\
&\qquad - \exp\left\{-\Lambda(t;0,1)\right\} \left[\widehat\Lambda(t;0,1)-\Lambda(t;0,1)\right] + o_p(m^{-1/2}) \\
&= \exp\left\{-\Lambda(t;0,0)\right\} \bigg[\sum_{n_1\in\{0,1\}}\int_0^t \widehat{w}_{n_1}(s^-;0)\overline{Y}_{n_1}^{-1}(s;0)d\overline{M}_{n_1}(s;0) \\
&\qquad\qquad\qquad + \int_0^t \{\widehat{w}_1(s^-;0)-w_1(s^-;0)\}\{d\Lambda_1(s;0)-d\Lambda_0(s;0)\}\bigg] + o_p(m^{-1/2}) \\
&\qquad - \exp\left\{-\Lambda(t;0,1)\right\} \bigg[\sum_{n_1\in\{0,1\}}\int_0^t \widehat{w}_{n_1}(s^-;0)\overline{Y}_{n_1}^{-1}(s;1)d\overline{M}_{n_1}(s;1) \\
&\qquad\qquad + \int_0^t \{\widehat{w}_1(s^-;0)-w_1(s^-;0)\}\{d\Lambda_1(s;1)-d\Lambda_0(s;1)\}\bigg] \\
&= \exp\left\{-\Lambda(t;0,0)\right\} \sum_{n_1\in\{0,1\}}\int_0^t \widehat{w}_{n_1}(s^-;0)\overline{Y}_{n_1}^{-1}(s;0)d\overline{M}_{n_1}(s;0) \\
&\qquad - \exp\left\{-\Lambda(t;0,1)\right\} \sum_{n_1\in\{0,1\}}\int_0^t \widehat{w}_{n_1}(s^-;0)\overline{Y}_{n_1}^{-1}(s;1)d\overline{M}_{n_1}(s;1) \\
&\qquad + \int_0^t \{\widehat{w}_1(s^-;0)-w_1(s^-;0)\} \bigg[\frac{d\Lambda_1(s;0)-d\Lambda_0(s;0)}{\exp\{\Lambda(t;0,0)\}} - \frac{d\Lambda_1(s;1)-d\Lambda_0(s;1)}{\exp\{\Lambda(t;0,1)\}}\bigg] \\
&\qquad + o_p(m^{-1/2}) \\
&= \sum_{z_2\in\{0,1\}} (-1)^{z_2} \exp\left\{-\Lambda(t;0,z_2)\right\} \sum_{n_1\in\{0,1\}}\int_0^t \frac{{w}_{n_1}(s^-;0)m^{-1/2}d\overline{M}_{n_1}(s;z_2)}{P(T_2 \wedge C \ge s, \widetilde{N}_1(s^-)=n_1, Z=z_2)} \\
&\qquad + \int_0^t m^{-1/2}X_1(s^-;0) \bigg[\frac{d\Lambda_1(s;0)-d\Lambda_0(s;0)}{\exp\{\Lambda(t;0,0)\}} - \frac{d\Lambda_1(s;1)-d\Lambda_0(s;1)}{\exp\{\Lambda(t;0,1)\}}\bigg] \\
&\qquad + o_p(m^{-1/2}).
\end{align*}
We see that $m^{1/2}\{\widehat\Delta_{\DE}(\cdot) - \Delta_{\DE}(\cdot)\}$ weakly converges to a sum of four independent martingales and a weighted Gaussian process. The asymptotic variance of $\widehat\Delta_{\DE}(t)$ is a sum of five terms corresponding to these processes.

\subsection{Estimated natural indirect effect}

The natural indirect effect (NIE) is estimated by
\begin{align*}
\widehat\Delta_{\IE}(t) &= \widehat{F}_{\prev}(t;1,1) - \widehat{F}_{\prev}(t;0,1) \\
&= \exp\left\{-\widehat\Lambda(t;0,1)\right\} - \exp\left\{-\widehat\Lambda(t;1,1)\right\}.
\end{align*}
To show its asymptotic law,
\begin{align*}
\widehat\Delta_{\IE} - \Delta_{\IE} &= \exp\left\{-\widehat\Lambda(t;0,1)\right\} - \exp\left\{-\Lambda(t;0,1)\right\} - \exp\left\{-\widehat\Lambda(t;1,1)\right\} + \exp\left\{-\Lambda(t;1,1)\right\} \\
&= \exp\left\{-\Lambda(t;0,1)\right\} \left[\widehat\Lambda(t;0,1)-\Lambda(t;0,1)\right] \\
&\qquad - \exp\left\{-\Lambda(t;1,1)\right\} \left[\widehat\Lambda(t;1,1)-\Lambda(t;1,1)\right] + o_p(m^{-1/2}) \\
&= \exp\left\{-\Lambda(t;0,1)\right\} \bigg[\sum_{n_1\in\{0,1\}}\int_0^t \widehat{w}_{n_1}(s^-;0)\overline{Y}_{n_1}^{-1}(s;1)d\overline{M}_{n_1}(s;1) \\
&\qquad\qquad +\int_0^t\{\widehat{w}_1(s^-;0)-w_1(s^-;0)\}\{d\Lambda_1(s;1)-d\Lambda_0(s;1)\}\bigg] \\
&\qquad - \exp\left\{-\Lambda(t;1,1)\right\} \bigg[\sum_{n_1\in\{0,1\}}\int_0^t \widehat{w}_{n_1}(s^-;1)\overline{Y}_{n_1}^{-1}(s;1)d\overline{M}_{n_1}(s;1) \\
&\qquad\qquad\qquad +\int_0^t\{\widehat{w}_1(s^-;1)-w_1(s^-;1)\}\{d\Lambda_1(s;1)-d\Lambda_0(s;1)\}\bigg] + o_p(m^{-1/2}) \\
&= \sum_{n_1\in\{0,1\}}\int_0^t \left[\frac{\widehat{w}_{n_1}(s^-;0)}{\exp\{\Lambda(t;0,1)\}} - \frac{\widehat{w}_{n_1}(s^-;1)}{\exp\{\Lambda(t;1,1)\}}\right] \overline{Y}_{n_1}^{-1}(s;1)d\overline{M}_{n_1}(s;1) \\
&\qquad + \int_0^t \left[\frac{\widehat{w}_1(s^-;0)-w_1(s^-;0)}{\exp\{\Lambda(t;0,1)\}} - \frac{\widehat{w}_1(s^-;1)-w_1(s^-;1)}{\exp\{\Lambda(t;1,1)\}}\right]\{d\Lambda_1(s;1)-d\Lambda_0(s;1)\} \\
&\qquad + o_p(m^{-1/2}) \\
&= \sum_{n_1\in\{0,1\}}\int_0^t \left[\frac{{w}_{n_1}(s^-;0)}{\exp\{\Lambda(t;0,1)\}} - \frac{{w}_{n_1}(s^-;1)}{\exp\{\Lambda(t;1,1)\}}\right] \frac{m^{-1/2}d\overline{M}_{n_1}(s;1)}{P(T_2 \wedge C \ge s, \widetilde{N}_1(s^-)=n_1, Z=1)} \\
&\qquad + \int_0^t m^{-1/2}\left[\frac{X_1(s^-;0)}{\exp\{\Lambda(t;0,1)\}} - \frac{X_1(s^-;1)}{\exp\{\Lambda(t;1,1)\}}\right]\{d\Lambda_1(s;1)-d\Lambda_0(s;1)\} \\
&\qquad + o_p(m^{-1/2}).
\end{align*}
We see that $m^{1/2}\{\widehat\Delta_{\IE}(\cdot) - \Delta_{\IE}(\cdot)\}$ weakly converges to a sum of two independent martingales and a weighted Gaussian process. The asymptotic variance of $\widehat\Delta_{\IE}(t)$ consists of a sum of three terms corresponding to these processes.

\section{Proof of Theorem 2}

\subsection{Identification}

Under Assumptions 2, 3, 4 and 8,
\begin{align*}
\frac{d}{dt}\Lambda_{*}(t; z_1, z_2) &:= \frac{1}{dt}P(d\widetilde{N}_1(t; z_1) = 1 \mid \widetilde{N}_1(t^-; z_1) = 0, \widetilde{N}_2(t^-; z_2, \widetilde{N}_1(\cdot;z_1)) = 0)) \\
&= \frac{1}{dt}P(d\widetilde{N}_1(t; z_1) = 1 \mid Z = z_1, \widetilde{N}_1(t^-; z_1) = 0, \widetilde{N}_2(t^-; z_2, \widetilde{N}_1(\cdot;z_1)) = 0) \\
&= \frac{1}{dt}P(d\widetilde{N}_1(t; z_1) = 1 \mid Z = z_1, \widetilde{N}_1(t^-; z_1) = 0, \widetilde{N}_2(t^-; z_1, \widetilde{N}_1(\cdot;z_1)) = 0) \\
&= \frac{1}{dt}P(d\widetilde{N}_1(t; z_1) = 1 \mid Z = z_1, \widetilde{N}_1(t^-; z_1) = 0, \widetilde{N}_2(t^-; z_1, \widetilde{N}_1(\cdot;z_1)) = 0, C \ge t) \\
&= \frac{1}{dt}P(d\widetilde{N}_1(t) = 1 \mid Z = z_1, \widetilde{N}_1(t^-) = 0, \widetilde{N}_2(t^-) = 0, C \ge t) \\
&= \frac{1}{dt}P(dN_1(t) = 1 \mid Z = z_1, Y_1(t) = 1, Y_2(t) = 1).
\end{align*}
This implies that $d\Lambda_*(t;z_1,z_1) = d\Lambda_*(t;z_1,z_2) := d\Lambda_*(t;z_1)$. Assumption 5 guarantees the conditional probability is estimable.

We partition the terminal event into a direct outcome event which does not have a history of non-terminal event and an indirect event which has a history of non-terminal event. The non-terminal event and direct terminal event can be considered as competing events with zero probability of ties (according to Remark 1), so
\begin{align*}
&\quad P(d\widetilde{N}_1(s;z_1) = 0, d\widetilde{N}_2(s;z_2,\widetilde{N}_1(\cdot;z_1)) = 0 \mid \widetilde{N}_1(s^-;z_1) = 0, \widetilde{N}_2(s;z_2,\widetilde{N}_1(\cdot;z_1)) = 0) \\
&= 1 - P(d\widetilde{N}_1(s;z_1) = 1 \mid \widetilde{N}_1(s^-;z_1) = 0, \widetilde{N}_2(s^-;z_2,\widetilde{N}_1(\cdot;z_1)) = 0) \\
&\qquad - P(d\widetilde{N}_2(s;z_2,\widetilde{N}_1(z_1)) = 1 \mid \widetilde{N}_1(s^-;z_1) = 0, \widetilde{N}_2(s^-;z_2,\widetilde{N}_1(\cdot;z_1)) = 0) \\
&= 1 - d\Lambda_{*}(s; z_1, z_2) - d\Lambda_0(s; z_1, z_2),
\end{align*}
and hence
\[
P(\widetilde{N}_1(s; z_1) = 0, \widetilde{N}_2(s; z_2, \widetilde{N}_1(\cdot;z_1)) = 0) = \exp\{-\Lambda_{*}(s; z_1, z_2) - \Lambda_0(s; z_1, z_2)\}.
\]
The cumulative incidence of the terminal event $F(t; z_1, z_2)$ consists of two subdistributions: a part without non-terminal event $F_0(t; z_1, z_2)$ and a part with non-terminal event $F_1(t; z_1, z_2)$, with
\begin{align*}
F_0(t; z_1, z_2) &= P(\widetilde{N}_1(t; z_1) = 0, \widetilde{N}_2(t; z_2, \widetilde{N}_1(\cdot;z_1)) = 1) \\
&= \int_0^t P(\widetilde{N}_1(s^-; z_1) = 0, \widetilde{N}_2(s^-; z_2, \widetilde{N}_1(\cdot;z_1)) = 0) \cdot d\Lambda_0(s; z_1, z_2) \\
&= \int_0^t \exp\{-\Lambda_{*}(s^-; z_1, z_2) - \Lambda_0(s^-; z_1, z_2)\} \cdot d\Lambda_0(s; z_1, z_2) \\
&= \int_0^t \exp\{-\Lambda_{*}(s^-; z_1) - \Lambda_0(s^-; z_1)\} \cdot d\Lambda_0(s; z_2), \\
F_1(t; z_1, z_2) &= P(\widetilde{N}_1(t; z_1) = 1, \widetilde{N}_2(t; z_2, \widetilde{N}_1(\cdot;z_1)) = 1) \\
&= \int_0^t P(d\widetilde{N}_1(s^-; z_1) = 1) \cdot P(\widetilde{N}_2(t; z_2, \widetilde{N}_1(\cdot;z_1)) = 1 \mid d\widetilde{N}_1(s^-; z_1) = 1) \\
&= \int_0^t P(\widetilde{N}_1(s^-; z_1) = 0, \widetilde{N}_2(s^-; z_2, \widetilde{N}_1(\cdot;z_1)) = 0) d\Lambda_{*}(s^-; z_1, z_2) \\
&\qquad \cdot P(\widetilde{N}_2(t; z_2, \widetilde{N}_1(\cdot;z_1)) = 1 \mid d\widetilde{N}_1(s; z_1) = 1, \widetilde{N}_2(s^-; z_2, \widetilde{N}_1(\cdot;z_1)) = 0) \\
&= \int_0^t \exp\{-\Lambda_{*}(s^-; z_1, z_2) - \Lambda_0(s^-; z_1, z_2)\} d\Lambda_{*}(s^-; z_1, z_2) \\
&\qquad \cdot [1 - \exp\{-\Lambda_1(t; z_1, z_2) + \Lambda_1(s^-; z_1, z_2)\}] \\
&= \int_0^t \exp\{-\Lambda_{*}(s^-; z_1) - \Lambda_0(s^-; z_2)\} d\Lambda_{*}(s; z_1) \cdot [1 - \exp\{-\Lambda_1(t; z_2) + \Lambda_1(s; z_2)\}].
\end{align*}
It is helpful to write the cumulative incidence of the non-terminal event
\begin{align*}
F_{*}(t;z_1,z_2) &= \int_0^t \exp\{-\Lambda_{*}(s^-; z_1) - \Lambda_0(s^-; z_2)\} \cdot d\Lambda_{*}(s; z_2),
\end{align*}
so the prevalence of the non-terminal event can be expressed by
$$w_1(t;z_1,z_2) = \frac{F_*(t;z_1,z_2) - F_1(t;z_1,z_2)}{1 - F_0(t;z_1,z_2) - F_1(t;z_1,z_2)}.$$

Finally, the counterfactual cumulative incidence $F(t; z_1, z_2)$ is identifiable, denoted by
\[
F_{\haz}(t; z_1, z_2) := F_0(t; z_1, z_2) + F_1(t; z_1, z_2).
\]

A consistent estimator for the counterfactual cumulative incidence is given by
\begin{align*}
\widehat{F}_{\haz}(t;z_1,z_2) &= \int_0^t \exp\{-\widehat\Lambda_{*}(s^-; z_1) - \widehat\Lambda_0(s^-; z_2)\} d\widehat\Lambda_0(s; z_2) \\
&\quad + \int_0^t \exp\{-\widehat\Lambda_{*}(s^-; z_1) - \widehat\Lambda_0(s^-; z_2)\} [1 - \exp\{-\widehat\Lambda_1(t; z_2) + \widehat\Lambda_1(s; z_2)\}] d\widehat\Lambda_{*}(s; z_1) \\
&= 1 - \exp\{-\widehat\Lambda_*(t;z_1)-\widehat\Lambda_0(t;z_2)\} \\
&\quad - \int_0^t \exp\{-\widehat\Lambda_*(s^-;z_1)-\widehat\Lambda_0(s^-;z_2)+\widehat\Lambda_1(s;z_2)-\widehat\Lambda_1(t;z_2)\}d\widehat\Lambda_*(s;z_1),
\end{align*}
where
\begin{align*}
d\widehat\Lambda_{n_1}(t; z_2) &= \frac{I\{\overline{Y}_{n_1}(t; z_2) > 0\}}{\overline{Y}_{n_1}(t; z_2)}d\overline{N}_{n_1}(t; z_2), \quad n_1 = 0, 1, \ z_2 = 0, 1, \\
d\widehat\Lambda_{*}(t; z_1) &= \frac{I\{\overline{Y}_{*}(t; z_1) > 0\}}{\overline{Y}_{*}(t; z_1)} d\overline{N}_{*}(t; z_1), \quad z_1 = 0, 1,
\end{align*}
with $\overline{N}_{n}(t; z)$ and $\overline{Y}_{n}(t; z)$ defined in Section 2.

\subsection{Estimated counterfactual cumulative incidence}

With a little abuse of notations, we view $\Lambda_{n}(t^-;z_1,z_2) = \Lambda_{n}(t;z_1,z_2)$ because the cumulative hazard is continuous, $n = *, 0, 1$. Let
\[
\overline{M}_{*}(t;z_2) = \int_0^t d\overline{N}_{*}(s;z_1) - \overline{Y}_{*}(s;z_1) d\Lambda_{*}(s;z_1)
\]
be the martingale with respect to $\overline{N}_{*}(s;z_2)$. Then
\begin{align*}
&\quad~\widehat{F}_{\haz}(t;z_1,z_2) - F_{\haz}(t;z_1,z_2) \\
&= 1 - \exp\{-\widehat\Lambda_*(t;z_1)-\widehat\Lambda_0(t;z_2)\} \\
&\qquad - \int_0^t \exp\{-\widehat\Lambda_*(s;z_1)-\widehat\Lambda_0(s;z_2)+\widehat\Lambda_1(s;z_2)-\widehat\Lambda_1(t;z_2)\}d\widehat\Lambda_*(s;z_1) \\
&\qquad - 1 + \exp\{-\Lambda_*(s;z_1)-\Lambda_0(s;z_2)\} \\
&\qquad + \int_0^t \exp\{-\Lambda_*(s;z_1)-\Lambda_0(s;z_2)+\Lambda_1(s;z_2)-\Lambda_1(t;z_2)\}d\Lambda_*(s;z_1) \\
&= \exp\{-\Lambda_*(t;z_1)-\Lambda_0(t;z_2)\} \int_0^t\left\{\frac{d\overline{M}_*(s;z_1)}{\overline{Y}_*(s;z_1)}+\frac{d\overline{M}_0(s;z_2)}{\overline{Y}_0(s;z_2)}\right\} \\
& \qquad + \int_0^t \exp\{-\Lambda_*(s;z_1)-\Lambda_0(s;z_2)+\Lambda_1(s;z_1)-\Lambda_1(t;z_2)\} \\
&\qquad\qquad \int_0^s\left\{\frac{d\overline{M}_*(u;z_1)}{\overline{Y}_*(u;z_1)}+\frac{d\overline{M}_0(u;z_2)}{\overline{Y}_0(u;z_2)}\right\} \\
& \qquad + \int_0^t \exp\{-\Lambda_*(s;z_1)-\Lambda_0(s;z_2)+\Lambda_1(s;z_1)-\Lambda_1(t;z_2)\} \int_s^t\frac{d\overline{M}_1(u;z_2)}{\overline{Y}_1(u;z_2)} \\
&\qquad - \int_0^t \exp\{-\Lambda_*(s;z_1)-\Lambda_0(s;z_2)+\Lambda_1(s;z_1)-\Lambda_1(t;z_2)\} \frac{d\overline{M}_*(s;z_1)}{\overline{Y}_*(s;z_1)} + o_p(m^{-1/2}) \\
&= \{1-F_0(t;z_1,z_2)-F_*(t;z_1,z_2)\} \int_0^t\left\{\frac{d\overline{M}_*(s;z_1)}{\overline{Y}_*(s;z_1)}+\frac{d\overline{M}_0(s;z_2)}{\overline{Y}_0(s;z_2)}\right\} \\
&\qquad + \int_0^t\int_s^t \exp\{-\Lambda_*(u;z_1)-\Lambda_0(u;z_2)+\Lambda_1(u;z_1)-\Lambda_1(t;z_2)\} d\Lambda_*(u;z_1) \\
&\qquad\qquad \left\{\frac{d\overline{M}_*(s;z_1)}{\overline{Y}_*(s;z_1)}+\frac{d\overline{M}_0(s;z_2)}{\overline{Y}_0(s;z_2)}\right\} \\
&\qquad + \int_0^t\int_0^s \exp\{-\Lambda_*(u;z_1)-\Lambda_0(u;z_2)+\Lambda_1(u;z_1)-\Lambda_1(t;z_2)\} d\Lambda_*(u;z_1)\frac{d\overline{M}_1(s;z_2)}{\overline{Y}_1(s;z_2)} \\
&\qquad - \int_0^t \exp\{-\Lambda_*(s;z_1)-\Lambda_0(s;z_2)+\Lambda_1(s;z_1)-\Lambda_1(t;z_2)\} \frac{d\overline{M}_*(s;z_1)}{\overline{Y}_*(s;z_1)} + o_p(m^{-1/2}) \\
&= \int_0^t [1-F_0(t;z_1,z_2)-F_1(t;z_1,z_2) \\
&\qquad\qquad -\{F_*(s;z_1,z_2)-F_1(s;z_1,z_2)\}\exp\{\Lambda_1(s;z_2)-\Lambda_1(t;z_2)\}] \frac{d\overline{M}_0(s;z_2)}{\overline{Y}_0(s;z_2)} \\
&\qquad\qquad +\{1-F_0(s;z_1,z_2)-F_*(s;z_1,z_2)\}\exp\{\Lambda_1(s;z_2)-\Lambda_1(t;z_2)\} \\
&\qquad\qquad +F_*(t;z_1,z_2)-F_1(t;z_1,z_2) \\
&\qquad\qquad -\{F_*(s;z_1,z_2)-F_1(s;z_1,z_2)\}\exp\{\Lambda_1(s;z_2)-\Lambda_1(t;z_2)\}] \frac{d\overline{M}_*(s;z_1)}{\overline{Y}_*(s;z_1)} \\
&\qquad + \int_0^t \{F_*(s;z_1,z_2)-F_1(s;z_1,z_2)\}\exp\{\Lambda_1(s;z_2)-\Lambda_1(t;z_2)\} \frac{d\overline{M}_1(s;z_2)}{\overline{Y}_1(s;z_2)} + o_p(m^{-1/2}) \\
&= \int_0^t  [1-F_{\haz}(t;z_1,z_2)-\{F_*(s;z_1,z_2)-F_1(s;z_1,z_2)\}\exp\{\Lambda_1(s;z_2)-\Lambda_1(t;z_2)\}] \frac{d\overline{M}_0(s;z_2)}{\overline{Y}_0(s;z_2)} \\
&\qquad + \int_0^t [1-F_{\haz}(t;z_1,z_2)-\{1-F_{\haz}(s;z_1,z_2)\}\exp\{\Lambda_1(s;z_2)-\Lambda_1(t;z_2)\}] \frac{d\overline{M}_*(s;z_1)}{\overline{Y}_*(s;z_1)} \\
&\qquad + \int_0^t \{F_*(s;z_1,z_2)-F_1(s;z_1,z_2)\}\exp\{\Lambda_1(s;z_2)-\Lambda_1(t;z_2)\} \frac{d\overline{M}_1(s;z_2)}{\overline{Y}_1(s;z_2)} + o_p(m^{-1/2}).
\end{align*}
The three terms above are independent martingales since they jump at different times. Note that $m^{-1}\overline{Y}_0(s;z) = m^{-1}\overline{Y}_*(s;z)$ converges to $P(T_1 \wedge T_2 \wedge C \ge s, Z=z)$, $m^{-1}\overline{Y}_1(s;z)$ converges to $P(T_1 < s, T_2 \wedge C \ge s, Z=z)$ by law of large numbers, and $m^{-1/2}\{\overline{M}_0(s;z), \overline{M}_*(s;z), \overline{M}_1(s;z)\}$ converge to Gaussian processes (also martingales) $\{M_0(s;z), M_*(s;z), M_1(s;z)\}$ with mean zero and variance $P(T_1 \wedge T_2 \wedge C \ge s, Z=z)\Lambda_0(s;z)$, $P(T_1 \wedge T_2 \wedge C \ge s, Z=z)\Lambda_*(s;z)$, $P(T_1 < s, T_2 \wedge C \ge s, Z=z)\Lambda_1(s;z)$ respectively by martingale central limit theorem. Applying the continuous mapping theorem,
\[
m^{1/2}\{\widehat{F}_{\haz}(t;z_1,z_2) - F_{\haz}(t;z_1,z_2)\} \xrightarrow{d} N\left\{0, \ \sigma_{\haz}^2(t;z_1,z_2)\right\},
\]
where
\begin{align*}
&\quad ~ \sigma_{\haz}^2(t;z_1,z_2) \\
&= \int_0^t \left[1-F_{\haz}(t;z_1,z_2)-\{F_*(s;z_1,z_2)-F_1(s;z_1,z_2)\}\frac{\exp\{\Lambda_1(s;z_2)}{\exp\{\Lambda_1(t;z_2)}\right]^2 \frac{d\Lambda_0(s;z_2)}{P(T_1 \wedge T_2 \wedge C \ge s, Z=z_2)} \\
&\qquad + \int_0^t \left[1-F_{\haz}(t;z_1,z_2)-\{1-F_{\haz}(s;z_1,z_2)\}\frac{\exp\{\Lambda_1(s;z_2)}{\exp\{\Lambda_1(t;z_2)}\right]^2 \frac{d\Lambda_*(s;z_1)}{P(T_1 \wedge T_2 \wedge C \ge s, Z=z_1)} \\
&\qquad + \int_0^t \left[\{F_*(s;z_1,z_2)-F_1(s;z_1,z_2)\}\frac{\exp\{\Lambda_1(s;z_2)}{\exp\{\Lambda_1(t;z_2)}\right]^2 \frac{d\Lambda_1(s;z_2)}{P(T_1 < s, T_2 \wedge C \ge s, Z=z_2)}.
\end{align*}

The hazard of non-terminal events is held unchanged when calculating the natural direct effect. Under this decomposition, NDE and NIE are estimated by $\widehat\Delta_{\DE}(t) = \widehat{F}_{\haz}(t;0,1) - \widehat{F}_{\haz}(t;0,0)$ and $\widehat\Delta_{\IE}(t) = \widehat{F}_{\haz}(t;1,1) - \widehat{F}_{\haz}(t;0,1)$, respectively. NDE measures the treatment effect on the cumulative incidence of terminal events via changing the hazards of terminal events while controlling the hazard of non-terminal events. NIE measures the treatment effect on the cumulative incidence of terminal events via changing the hazard of terminal events while controlling the hazards of terminal events.

\subsection{Estimated natural direct effect}

The natural direct effect (NDE) is estimated by
\begin{align*}
\widehat\Delta_{\DE}(t) = \widehat{F}_{\haz}(t;0,1) - \widehat{F}_{\haz}(t;0,0).
\end{align*}
To show its asymptotic law,
\begin{align*}
&\quad ~ \widehat\Delta_{\DE}(t) - \Delta_{\DE}(t) \\
&= \int_0^t  [1-F_{\haz}(t;0,1)-\{F_*(s;0,1)-F_1(s;0,1)\}\exp\{\Lambda_1(s;1)-\Lambda_1(t;1)\}] \frac{d\overline{M}_0(s;1)}{\overline{Y}_0(s;1)} \\
&\qquad - \int_0^t  [1-F_{\haz}(t;0,0)-\{F_*(s;0,0)-F_1(s;0,0)\}\exp\{\Lambda_1(s;0)-\Lambda_1(t;0)\}] \frac{d\overline{M}_0(s;0)}{\overline{Y}_0(s;0)} \\
&\qquad + \int_0^t [1-F_{\haz}(t;0,1)-\{1-F_{\haz}(s;0,1)\}\exp\{\Lambda_1(s;1)-\Lambda_1(t;1)\} \\
&\qquad\qquad -1+F_{\haz}(t;0,0)+\{1-F_{\haz}(s;0,0)\}\exp\{\Lambda_1(s;0)-\Lambda_1(t;0)\}] \frac{d\overline{M}_*(s;0)}{\overline{Y}_*(s;0)} \\
&\qquad + \int_0^t \{F_*(s;0,1)-F_1(s;0,1)\}\exp\{\Lambda_1(s;1)-\Lambda_1(t;1)\} \frac{d\overline{M}_1(s;1)}{\overline{Y}_1(s;1)} \\
&\qquad - \int_0^t \{F_*(s;0,0)-F_1(s;0,0)\}\exp\{\Lambda_1(s;0)-\Lambda_1(t;0)\} \frac{d\overline{M}_1(s;0)}{\overline{Y}_1(s;0)} + o_p(m^{-1/2}).
\end{align*}
We see that $m^{1/2}\{\widehat\Delta_{\DE}(\cdot) - \Delta_{\DE}(\cdot)\}$ weakly converges to a Gaussian process comprised of five independent martingales. The calculation of the asymptotic variance is trivial using the martingale theory.

\subsection{Estimated natural indirect effect}

The natural indirect effect (NIE) is estimated by
\begin{align*}
\widehat\Delta_{\IE}(t) = \widehat{F}_{\haz}(t;1,1) - \widehat{F}_{\haz}(t;0,1).
\end{align*}
To show its asymptotic law,
\begin{align*}
&\quad ~ \widehat\Delta_{\IE}(t) - \Delta_{\IE}(t) \\
&= \int_0^t  [1-F_{\haz}(t;1,1)-\{F_*(s;1,1)-F_1(s;1,1)\}\exp\{\Lambda_1(s;1)-\Lambda_1(t;1)\} \\
&\qquad\qquad -1+F_{\haz}(t;0,1)+\{F_*(s;0,1)-F_1(s;0,1)\}\exp\{\Lambda_1(s;1)-\Lambda_1(t;1)\}] \frac{d\overline{M}_0(s;1)}{\overline{Y}_0(s;1)} \\
&\qquad + \int_0^t [1-F_{\haz}(t;1,1)-\{1-F_{\haz}(s;1,1)\}\exp\{\Lambda_1(s;1)-\Lambda_1(t;1)\}] \frac{d\overline{M}_*(s;1)}{\overline{Y}_*(s;1)} \\
&\qquad - \int_0^t [1-F_{\haz}(t;0,1)-\{1-F_{\haz}(s;0,1)\}\exp\{\Lambda_1(s;1)-\Lambda_1(t;1)\}] \frac{d\overline{M}_*(s;0)}{\overline{Y}_*(s;0)} \\
&\qquad + \int_0^t [\{F_*(s;1,1)-F_1(s;1,1)\}\exp\{\Lambda_1(s;1)-\Lambda_1(t;1)\} \\
&\qquad\qquad -\{F_*(s;0,1)-F_1(s;0,1)\}\exp\{\Lambda_1(s;1)-\Lambda_1(t;1)\}] \frac{d\overline{M}_1(s;1)}{\overline{Y}_1(s;1)} + o_p(m^{-1/2}).
\end{align*}
We see that $m^{1/2}\{\widehat\Delta_{\IE}(\cdot) - \Delta_{\IE}(\cdot)\}$ weakly converges to a Gaussian process comprised of four independent martingales. The calculation of the asymptotic variance is trivial using the martingale theory.

\section{Numerical Studies} \label{sec:simulation}

To demonstrate the subtle difference between these two decompositions, we conduct a series of simple simulation studies in this section.
Suppose the underlying single-world data generating process satisfies Assumptions 1--6, with
\begin{align*}
\frac{d}{dt}\Lambda_{0}(t;z) &= (0.10 - 0.05a z)t, \\
\frac{d}{dt}\Lambda_{*}(t;z) &= (0.08 - 0.04b z)t, \\
\frac{d}{dt}\Lambda_{1}(t;z) &= (0.30 - 0.10c z)t.
\end{align*}
Each individual is assigned to treatment or control with equal probability $P(Z=1)=0.5$. The censoring distribution is uniform in $[6,10]$, independent of treatment assginments and potential failure times.

Then we consider three settings: (1) $a=1$, $b=c=0$, the hazard of the direct outcome event varies; (2) $b=1$, $a=c=0$, the hazard of the non-terminal event varies (which leads to change of the prevalence); (3) $c=1$, $a=b=0$, the hazard of the indirect outcome event varies.
Suppose the sample size $m = 500$.

\begin{figure}
\centering
\includegraphics[width=0.95\textwidth]{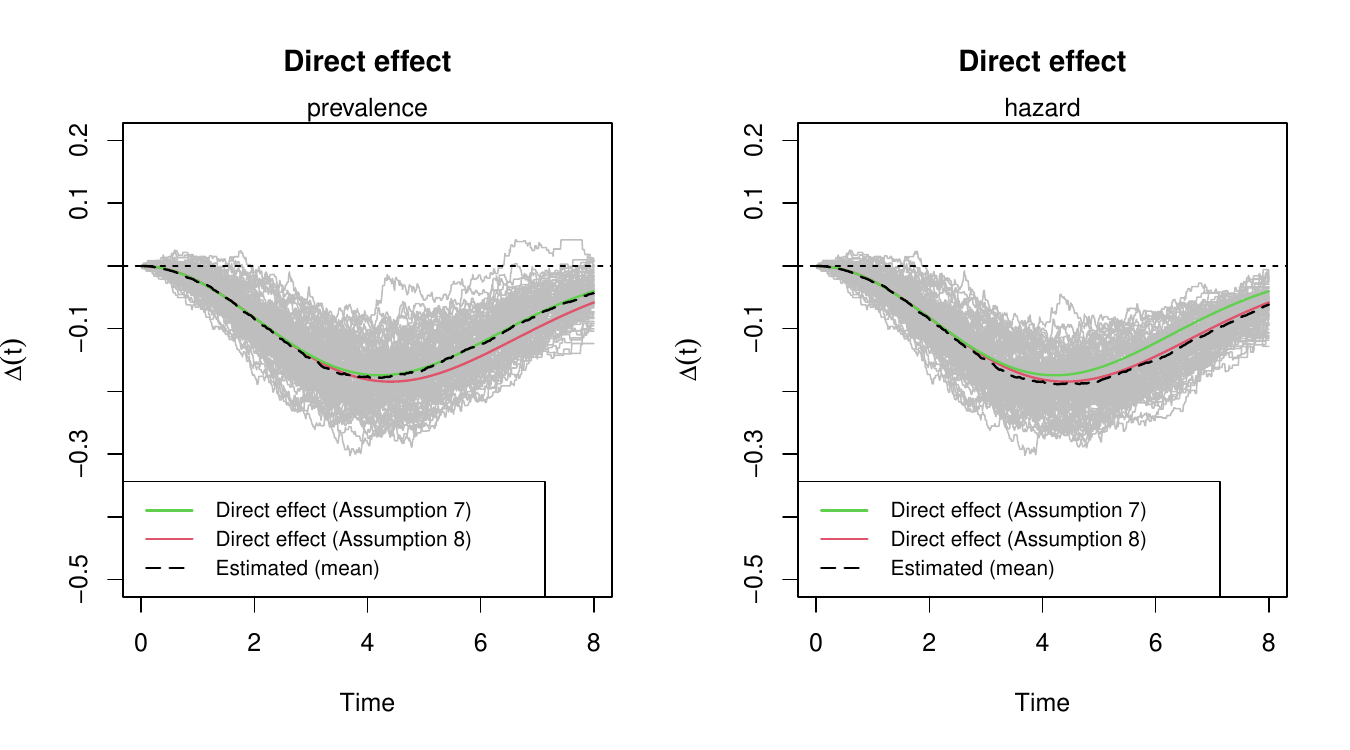}
\includegraphics[width=0.95\textwidth]{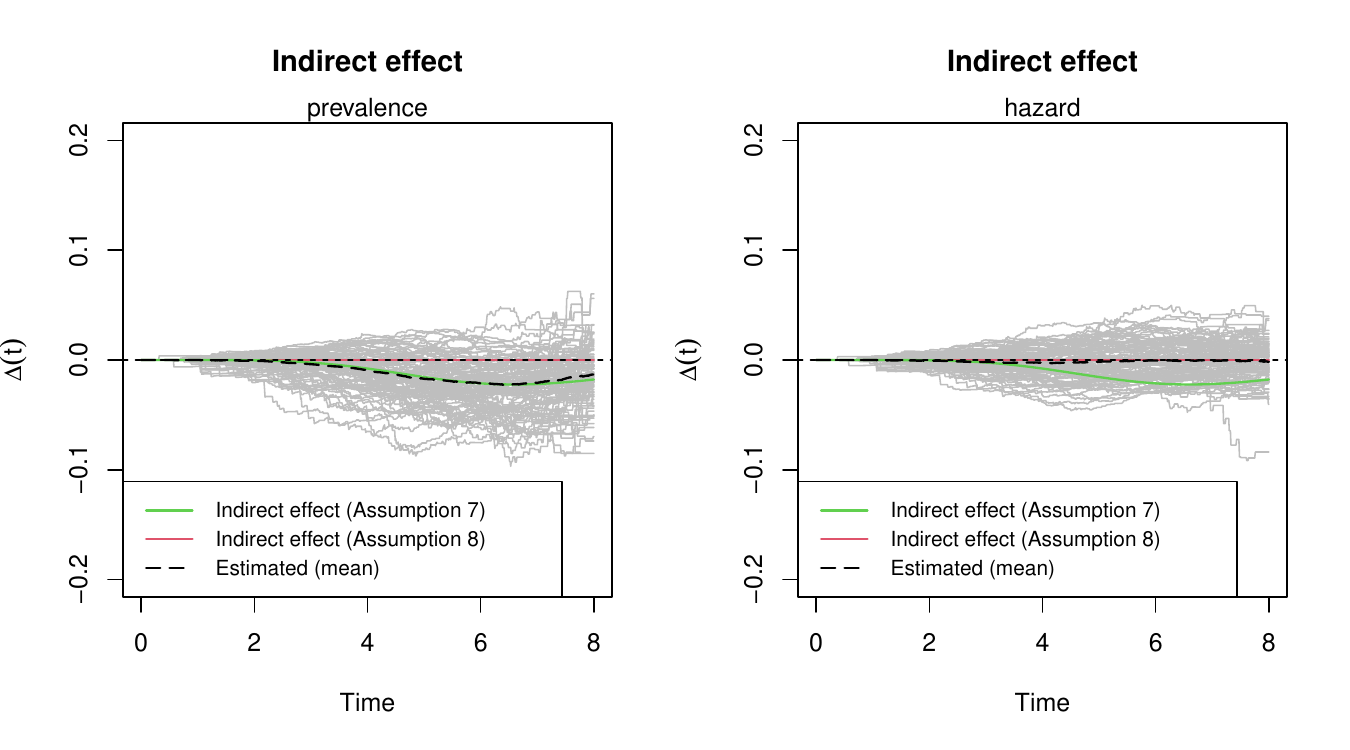} \\
\caption{Setting 1. True treatment effect under Assumption 7 (green, solid), true treatment effect under Assumption 8 (red, solid), estimated treatment effects (gray, solid) and estimated mean treatment effects (black, dashed). Left: Decomposition 1; Right: Decomposition 2.} \label{fig_simu1}
\end{figure}

\begin{figure}
\centering
\includegraphics[width=0.95\textwidth]{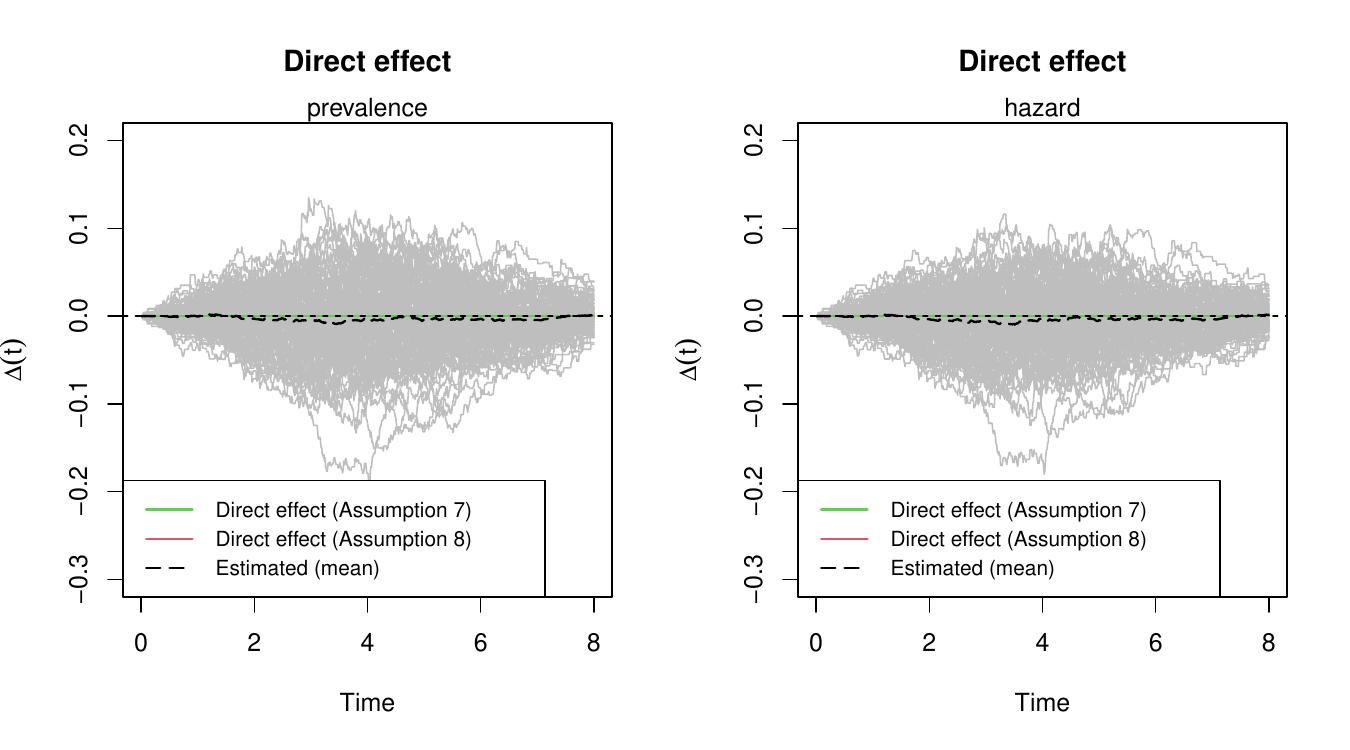}
\includegraphics[width=0.95\textwidth]{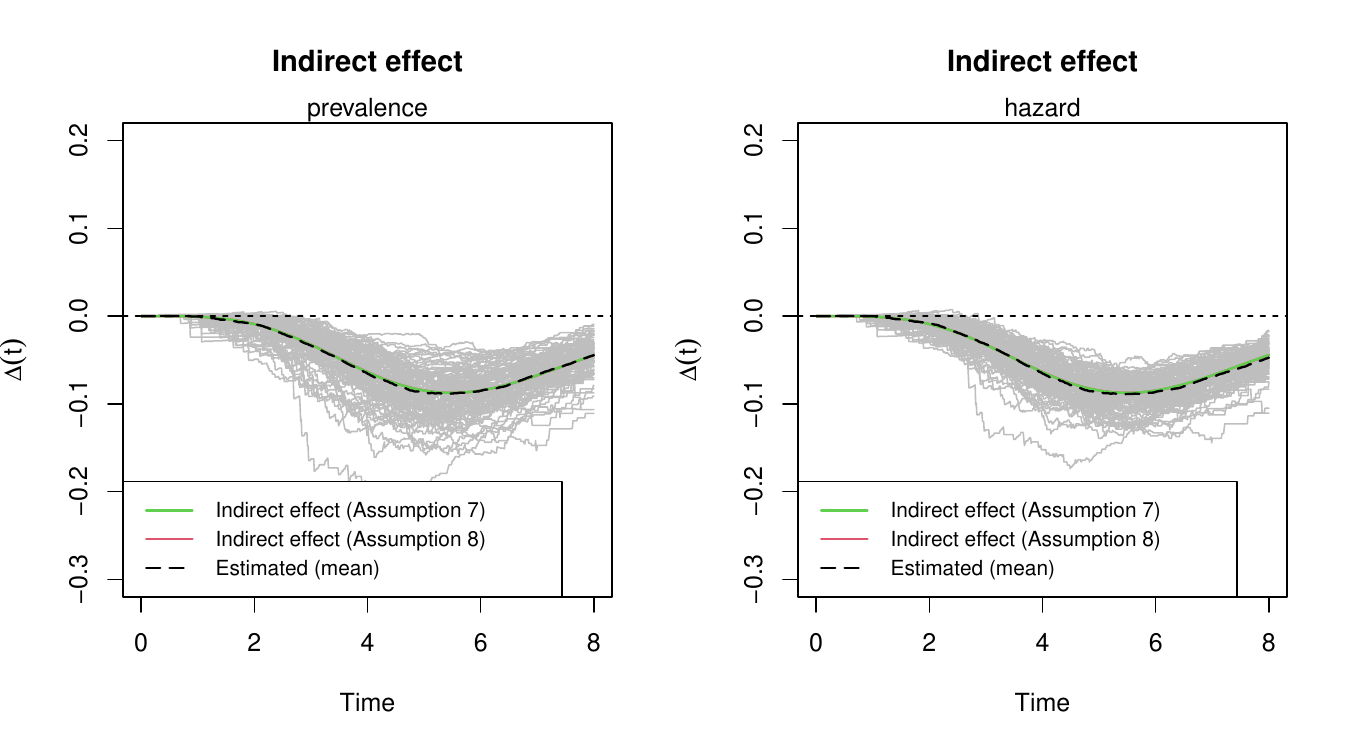} \\
\caption{Setting 2. True treatment effect under Assumption 7 (green, solid), true treatment effect under Assumption 8 (red, solid), estimated treatment effects (gray, solid) and estimated mean treatment effects (black, dashed). Red line and green line coincide. Left: Decomposition 1; Right: Decomposition 2.} \label{fig_simu2}
\end{figure}

\begin{figure}
\centering
\includegraphics[width=0.95\textwidth]{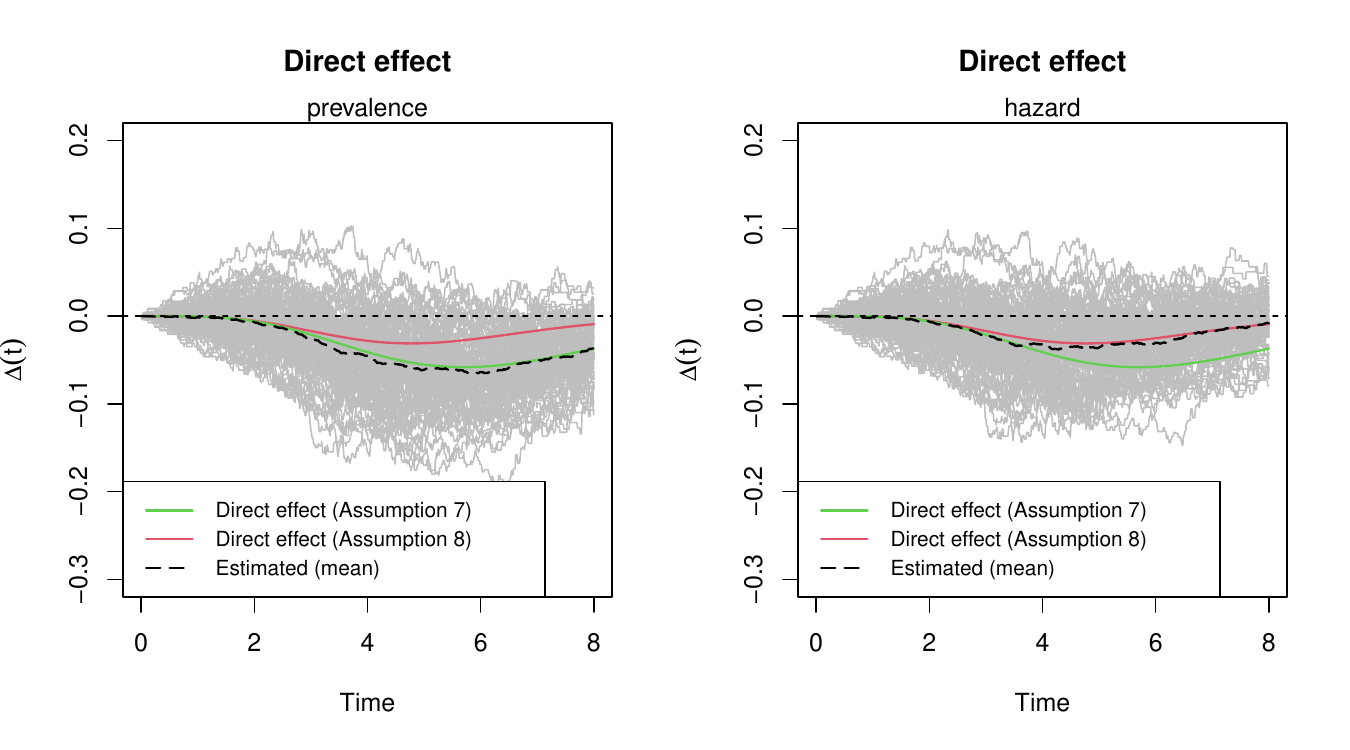}
\includegraphics[width=0.95\textwidth]{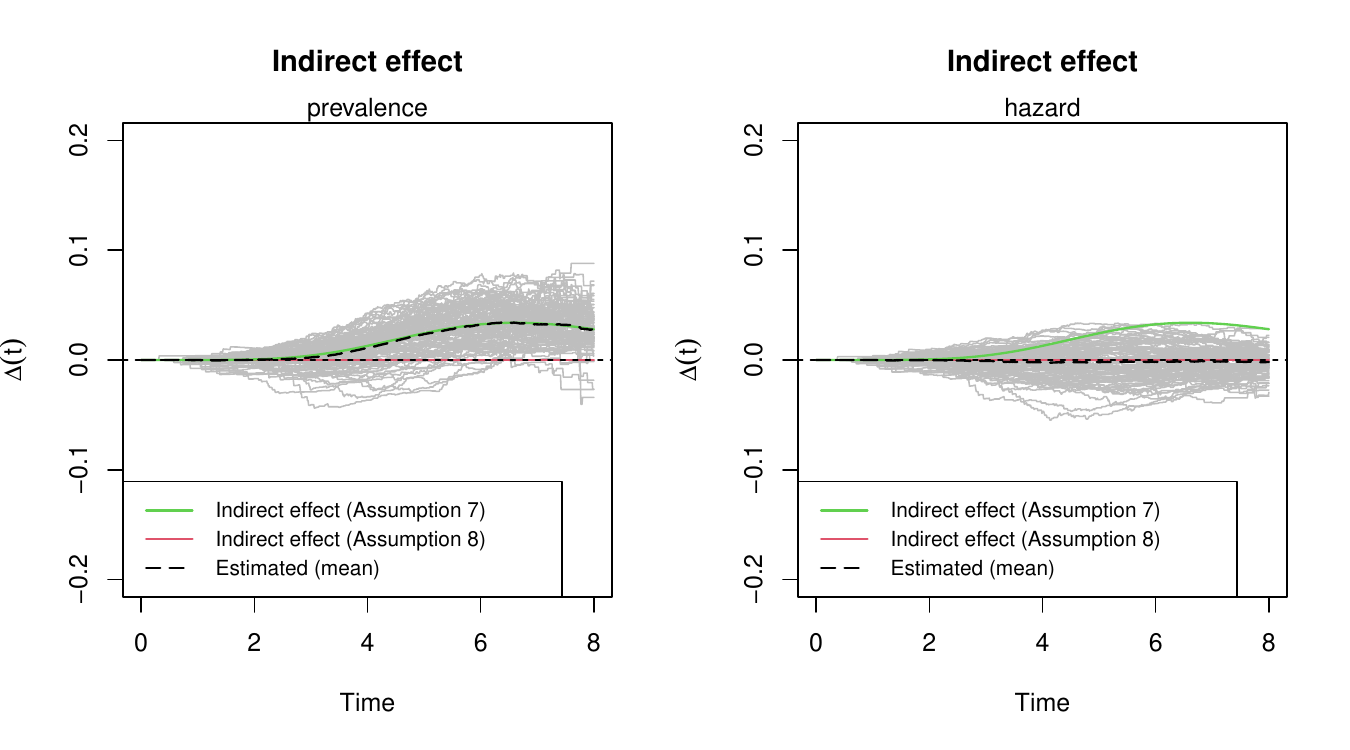} \\
\caption{Setting 3. True treatment effect under Assumption 7 (green, solid), true treatment effect under Assumption 8 (red, solid), estimated treatment effects (gray, solid) and estimated mean treatment effects (black, dashed). Left: Decomposition 1; Right: Decomposition 2.} \label{fig_simu3}
\end{figure}

We display the estimated natural direct and indirect effects under both decompositions in Figures \ref{fig_simu1}--\ref{fig_simu3}, corresponding to Settings 1--3, respectively, among 100 independently generated datasets. The left panels are estimates by Decomposition 1, and the right panels are estimates by Decomposition 2. In Setting 2, both assumptions lead to the same natural direct or indirect effect, because the hazards of the terminal event remain unchanged. It is interesting to observe that these decompositions give different estimates in Setting 1 and Setting 3, indicating that varying the hazards of the terminal event may contribute to natural indirect effects under Decomposition 1. This is because that the prevalence of non-terminal events relies on the hazards of the terminal event through modifying the population of at-risk individuals for the terminal event.

\begin{table}
\centering
\caption{Standard error (SE) and standard deviation (SD) at specific timepoints by estimation under two decompositions} \label{sesd}
\begin{tabular}{cccccccccccc}
\toprule
\multicolumn{2}{c}{Setting} & SE/SD & \multicolumn{4}{c}{Decomposition 1} & \multicolumn{4}{c}{Decomposition 2} \\ \cmidrule(lr){4-7} \cmidrule(lr){8-11}
\multicolumn{2}{c}{Estimand} & Type of SE & 2 & 4 & 6 & 8 & 2 & 4 & 6 & 8 \\
   \midrule
\multicolumn{11}{l}{Assumption 7 is correct} \\
1 & DE & Asymptotic SE & 0.032 & 0.045 & 0.037 & 0.031 & 0.031 & 0.043 & 0.033 & 0.023 \\
  &    & Bootstrap SE & 0.031 & 0.046 & 0.038 & 0.028 & 0.031 & 0.043 & 0.034 & 0.027 \\
  &    & SD & 0.032 & 0.046 & 0.039 & 0.028 & 0.031 & 0.043 & 0.034 & 0.026 \\
1 & IE & Asymptotic SE & 0.009 & 0.018 & 0.025 & 0.033 & 0.004 & 0.015 & 0.019 & 0.015 \\
  &    & Bootstrap SE & 0.006 & 0.019 & 0.027 & 0.025 & 0.005 & 0.015 & 0.019 & 0.020 \\
  &    & SD & 0.005 & 0.019 & 0.027 & 0.025 & 0.004 & 0.014 & 0.019 & 0.019 \\
2 & DE & Asymptotic SE & 0.030 & 0.049 & 0.037 & 0.021 & 0.029 & 0.043 & 0.030 & 0.015 \\
  &    & Bootstrap SE & 0.030 & 0.048 & 0.036 & 0.019 & 0.029 & 0.043 & 0.031 & 0.017 \\
  &    & SD & 0.030 & 0.049 & 0.036 & 0.018 & 0.029 & 0.044 & 0.030 & 0.016 \\
2 & IE & Asymptotic SE & 0.012 & 0.023 & 0.023 & 0.024 & 0.008 & 0.020 & 0.019 & 0.014 \\
  &    & Bootstrap SE & 0.010 & 0.026 & 0.026 & 0.019 & 0.008 & 0.020 & 0.020 & 0.016 \\
  &    & SD & 0.010 & 0.025 & 0.026 & 0.018 & 0.008 & 0.019 & 0.019 & 0.015 \\
3 & DE & Asymptotic SE & 0.028 & 0.044 & 0.040 & 0.034 & 0.027 & 0.041 & 0.034 & 0.023 \\
  &    & Bootstrap SE & 0.028 & 0.043 & 0.039 & 0.031 & 0.027 & 0.041 & 0.035 & 0.026 \\
  &    & SD & 0.027 & 0.043 & 0.038 & 0.033 & 0.026 & 0.041 & 0.034 & 0.026 \\
3 & IE & Asymptotic SE & 0.009 & 0.017 & 0.021 & 0.028 & 0.004 & 0.014 & 0.017 & 0.012 \\
  &    & Bootstrap SE & 0.006 & 0.016 & 0.021 & 0.021 & 0.005 & 0.014 & 0.017 & 0.014 \\
  &    & SD & 0.005 & 0.016 & 0.021 & 0.020 & 0.004 &0.014 & 0.017 & 0.013 \\
\midrule
\multicolumn{11}{l}{Assumption 8 is correct} \\
1 & DE & Asymptotic SE & 0.032 & 0.045 & 0.037 & 0.031 & 0.031 & 0.043 & 0.033 & 0.024 \\
  &    & Bootstrap SE & 0.032 & 0.046 & 0.038 & 0.028 & 0.031 & 0.043 & 0.034 & 0.027 \\
  &    & SD & 0.032 & 0.046 & 0.039 & 0.028 & 0.031 & 0.043 & 0.034 & 0.026 \\
1 & IE & Asymptotic SE & 0.009 & 0.018 & 0.025 & 0.033 & 0.004 & 0.015 & 0.019 & 0.015 \\
 &  & Bootstrap SE & 0.006 & 0.020 & 0.027 & 0.025 & 0.005 & 0.015 & 0.019 & 0.019 \\
 &  & SD & 0.005 & 0.019 & 0.027 & 0.025 & 0.004 & 0.014 & 0.019 & 0.018 \\
2 & DE & Asymptotic SE & 0.030 & 0.049 & 0.037 & 0.021 & 0.029 & 0.044 & 0.031 & 0.017 \\
 &  & Bootstrap SE & 0.030 & 0.049 & 0.036 & 0.019 & 0.029 & 0.044 & 0.031 & 0.017 \\
 &  & SD & 0.030 & 0.049 & 0.036 & 0.018 & 0.029 & 0.044 & 0.030 & 0.016 \\
2 & IE & Asymptotic SE & 0.012 & 0.023 & 0.023 & 0.024 & 0.008 & 0.019 & 0.019 & 0.014 \\
 &  & Bootstrap SE & 0.010 & 0.026 & 0.026 & 0.019 & 0.008 & 0.020 & 0.020 & 0.016 \\
 &  & SD & 0.010 & 0.025 & 0.026 & 0.018 & 0.008 & 0.019 & 0.019 & 0.015 \\
3 & DE & Asymptotic SE & 0.028 & 0.044 & 0.040 & 0.034 & 0.027 & 0.041 & 0.034 & 0.024 \\
 &  & Bootstrap SE & 0.028 & 0.043 & 0.039 & 0.031 & 0.027 & 0.041 & 0.035 & 0.026 \\
 &  & SD & 0.027 & 0.043 & 0.038 & 0.033 & 0.026 & 0.041 & 0.034 & 0.027 \\
3 & IE & Asymptotic SE & 0.009 & 0.017 & 0.021 & 0.028 & 0.004 & 0.014 & 0.017 & 0.012 \\
 &  & Bootstrap SE & 0.006 & 0.016 & 0.021 & 0.022 & 0.005 & 0.014 & 0.017 & 0.014 \\
 &  & SD & 0.005 & 0.016 & 0.021 & 0.020 & 0.004 & 0.014 & 0.017 & 0.013 \\
 \bottomrule
\end{tabular}
\end{table}

\begin{table}
\centering
\caption{Coverage of 95\% asymptotic and bootstrap (200 resamplings) pointwise confidence intervals at specific timepoints by estimation under two decompositions} \label{cicv}
\begin{tabular}{ccccccccccc}
\toprule
\multicolumn{2}{c}{Setting} & Type of & \multicolumn{4}{c}{Decomposition 1} & \multicolumn{4}{c}{Decomposition 2} \\ \cmidrule(lr){4-7} \cmidrule(lr){8-11}
\multicolumn{2}{c}{Estimand} & Conf. Int. & 2 & 4 & 6 & 8 & 2 & 4 & 6 & 8 \\
   \midrule
\multicolumn{11}{l}{Assumption 7 is correct} \\
1 & DE & Asymptotic & 0.949 & 0.945 & 0.927 & 0.954 & 0.947 & 0.936 & 0.895 & 0.856 \\
 &  & Bootstrap & 0.945 & 0.951 & 0.954 & 0.930 & 0.946 & 0.937 & 0.907 & 0.919 \\
1 & IE & Asymptotic & 1.000 & 0.965 & 0.936 & 0.992 & 0.993 & 0.930 & 0.785 & 0.706 \\
 &  & Bootstrap & 0.996 & 0.961 & 0.938 & 0.940 & 0.994 & 0.927 & 0.806 & 0.862 \\
2 & DE & Asymptotic & 0.948 & 0.947 & 0.947 & 0.984 & 0.939 & 0.949 & 0.948 & 0.947 \\
 &  & Bootstrap & 0.950 & 0.945 & 0.941 & 0.951 & 0.948 & 0.939 & 0.953 & 0.967 \\
2 & IE & Asymptotic & 0.986 & 0.909 & 0.917 & 0.995 & 0.832 & 0.937 & 0.945 & 0.926 \\
 &  & Bootstrap & 0.837 & 0.928 & 0.929 & 0.939 & 0.857 & 0.919 & 0.940 & 0.949 \\
3 & DE & Asymptotic & 0.964 & 0.962 & 0.968 & 0.958 & 0.964 & 0.947 & 0.847 & 0.744 \\
 &  & Bootstrap & 0.949 & 0.950 & 0.956 & 0.952 & 0.946 & 0.931 & 0.838 & 0.805 \\
3 & IE & Asymptotic & 1.000 & 0.976 & 0.967 & 0.994 & 0.995 & 0.852 & 0.487 & 0.334 \\
 &  & Bootstrap & 0.996 & 0.963 & 0.965 & 0.963 & 0.997 & 0.839 & 0.502 & 0.429 \\
\midrule
\multicolumn{11}{l}{Assumption 8 is correct} \\
1 & DE & Asymptotic & 0.949 & 0.946 & 0.909 & 0.895 & 0.947 & 0.949 & 0.940 & 0.931 \\
 &  & Bootstrap & 0.950 & 0.952 & 0.898 & 0.879 & 0.952 & 0.948 & 0.951 & 0.947 \\
1 & IE & Asymptotic & 1.000 & 0.938 & 0.883 & 0.973 & 0.994 & 0.965 & 0.952 & 0.943 \\
 &  & Bootstrap & 0.999 & 0.965 & 0.907 & 0.916 & 0.998 & 0.964 & 0.967 & 0.977 \\
2 & DE & Asymptotic & 0.948 & 0.947 & 0.947 & 0.984 & 0.943 & 0.949 & 0.950 & 0.962 \\
 &  & Bootstrap & 0.952 & 0.957 & 0.951 & 0.970 & 0.949 & 0.948 & 0.958 & 0.966 \\
2 & IE & Asymptotic & 0.987 & 0.910 & 0.915 & 0.995 & 0.830 & 0.938 & 0.945 & 0.938 \\
 &  & Bootstrap & 0.853 & 0.943 & 0.952 & 0.951 & 0.871 & 0.943 & 0.961 & 0.949 \\
3 & DE & Asymptotic & 0.962 & 0.943 & 0.877 & 0.867 & 0.965 & 0.955 & 0.958 & 0.921 \\
 &  & Bootstrap & 0.943 & 0.926 & 0.875 & 0.839 & 0.944 & 0.940 & 0.944 & 0.956 \\
3 & IE & Asymptotic & 1.000 & 0.934 & 0.691 & 0.928 & 0.997 & 0.967 & 0.952 & 0.961 \\
 &  & Bootstrap & 0.998 & 0.849 & 0.613 & 0.750 & 0.999 & 0.973 & 0.962 & 0.985 \\
 \bottomrule
\end{tabular}
\end{table}

Table \ref{sesd} shows the standard error (by asymptotic formula and bootstrap) and standard deviation of the estimates at some timepoints among 1000 independently generated datasets. Table \ref{cicv} shows the coverage rates of pointwise 95\% asymptotic confidence intervals and bootstrap (200 resamplings) confidence intervals. Decomposition 1 adopts Assumption 7 and Decomposition 2 adopts Assumption 8. In Setting 2, both decompositions should be correct because the hazards of the terminal event remain identical. Decomposition 2 generally yields smaller standard errors and shorter confidence intervals than Decomposition 1. The confidence intervals by correctly choosing Assumption 7 or 8 show good coverage rates in the middle part. The confidence intervals do not have perfect coverage rates at the head (when the failure time is small) or at the tail (when the failure time is large). At the head, both the treatment effect and the variance of treatment effect estimator are small, especially for indirect effects (see Figures \ref{fig_simu1}--\ref{fig_simu3}). Since the convergence of the estimator to a normal distribution is asymptotic with a higher-order bias of $o_p(m^{-1/2})$, the relative error of the estimated variance could be large compared with the true variance, resulting in a skewed distribution of estimates and unsatisfactory coverage. At the tail, there are few units still at risk, so the convergence is slow. Bootstrap confidence intervals show slightly better coverage rates at the tail.

\section{Application to Real Data} \label{sec:application}

\subsection{Hepatitis B on mortality mediated by liver cancer}

Hepatitis B causes serious public health burden worldwide \citep{chen2018global}. We are interested in the mechanism of hepatitis B on overall mortality in order to prevent the negative consequences of hepatitis B. The data came from a prospective cohort study designed to study the natural history of chronic hepatitis B in the development of liver cancer \citep{huang2011lifetime}. The sample in this study is restricted to 4954 male participants with age at cohort entry less than 50 years and without history of alcohol consumption. This dataset has been analyzed in \citet{huang2021causal}. Here we modify the target estimand to the difference in cumulative incidences. The treatment is hepatitis B infection (1 for positive and 0 for negative), the non-terminal event is occurrence of liver cancer, and the terminal event is death.

\begin{figure}
\centering
\includegraphics[width=0.95\textwidth]{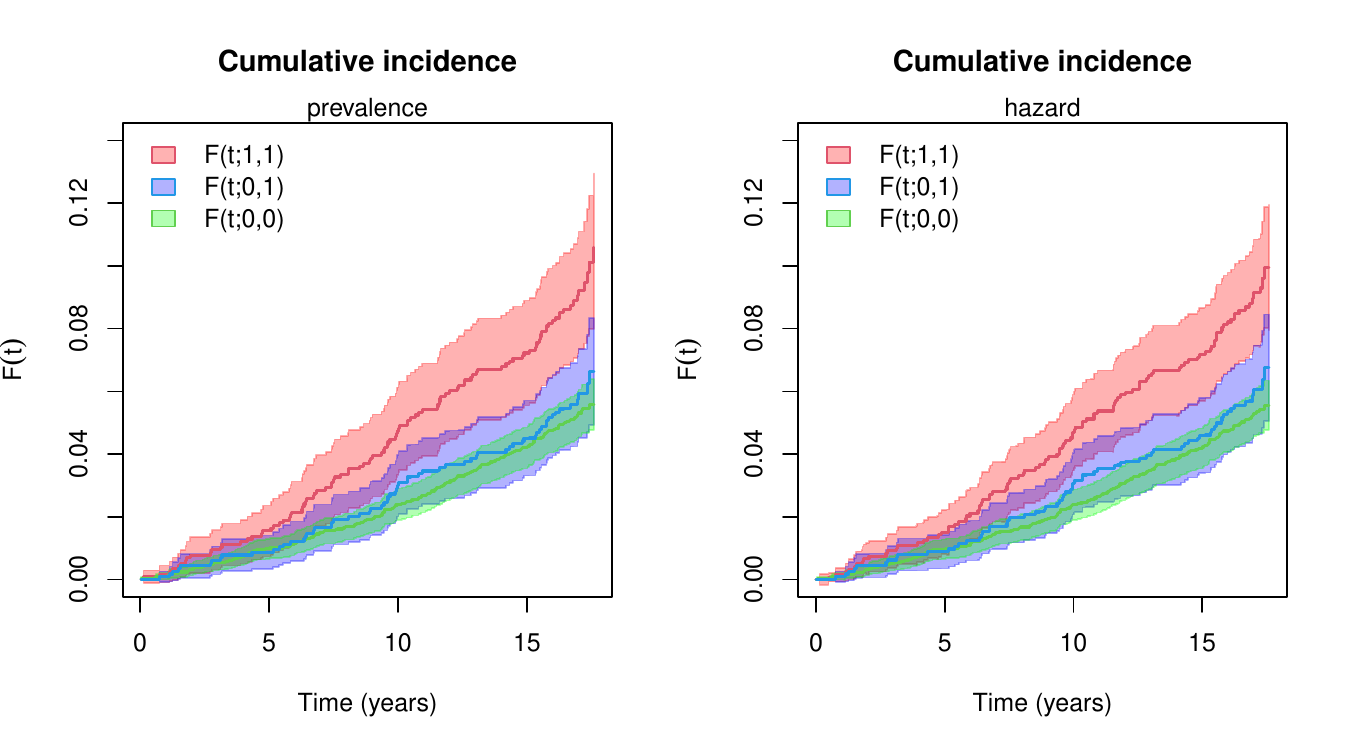} \\
\includegraphics[width=0.95\textwidth]{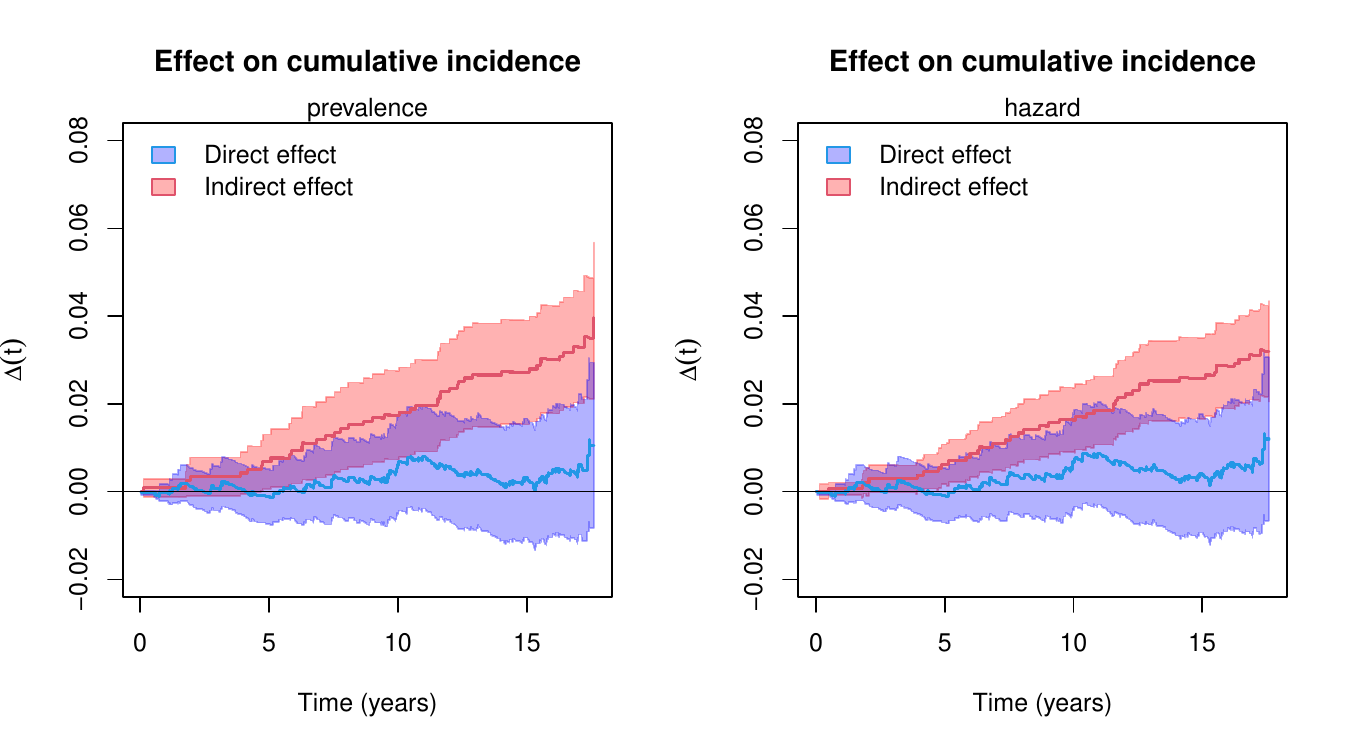}
\caption{Estimated counterfactual cumulative incidences and treatment effects with 95\% asymptotic confidence intervals for the hepatitis B data. Left: Decomposition 1 (controlling the prevalence); right: Decomposition 2 (controlling the hazard).} \label{fig_reveal}
\end{figure}

Figure \ref{fig_reveal} shows the counterfactual cumulative incidences of mortality and decomposed treatment effects with 95\% asymptotic confidence intervals (CIs). The left two figures decompose the total effect according to Decomposition 1 (controlling the prevalence), and the right two figures decompose the total effect according to Decomposition 2 (controlling the hazard). The estimated cumulative incidences, natural direct effects and natural indirect effects seem similar under both decompositions. The 95\% CI for NDE covers zero while the 95\% CI for NIE remains positive with time going on. This result leads to the conclusion that hepatitis B increases the risk of mortality mainly by increasing the risk of liver cancer.

\begin{figure}
\centering
\includegraphics[width=0.95\textwidth]{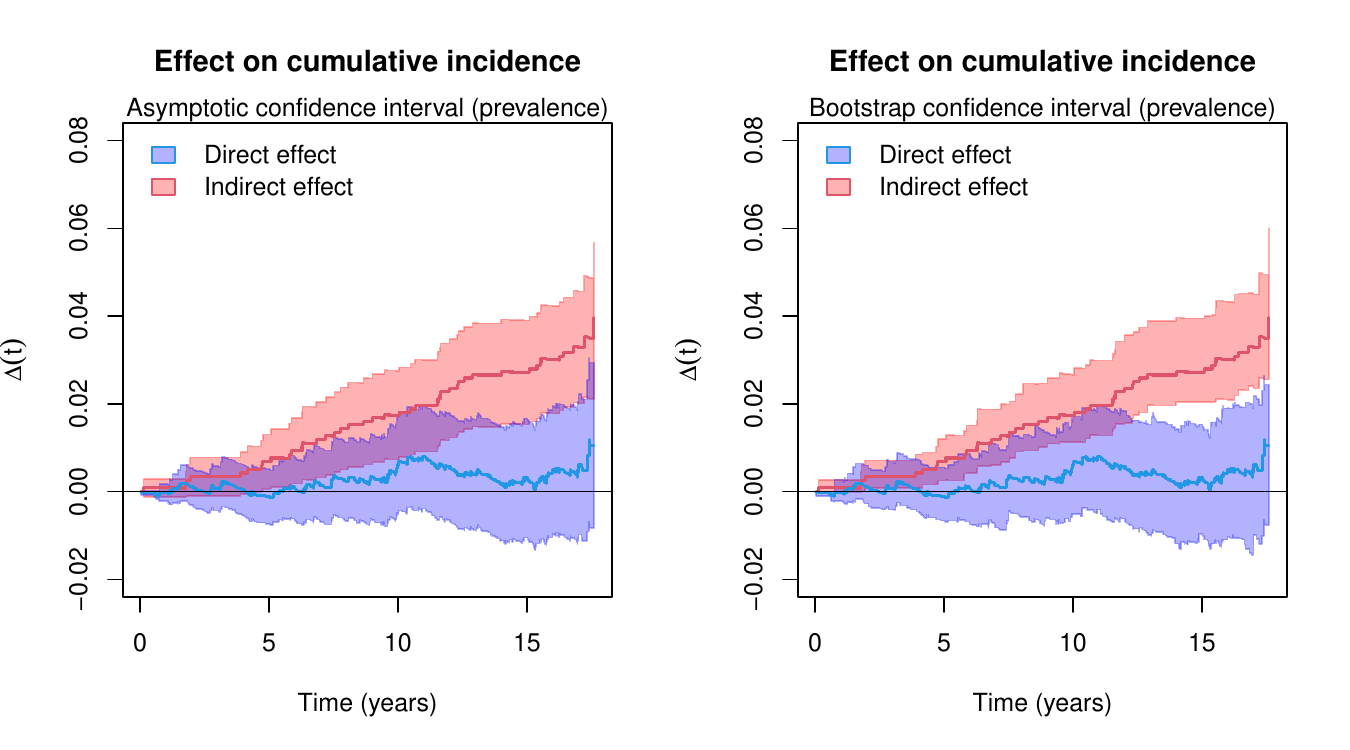} \\
\includegraphics[width=0.95\textwidth]{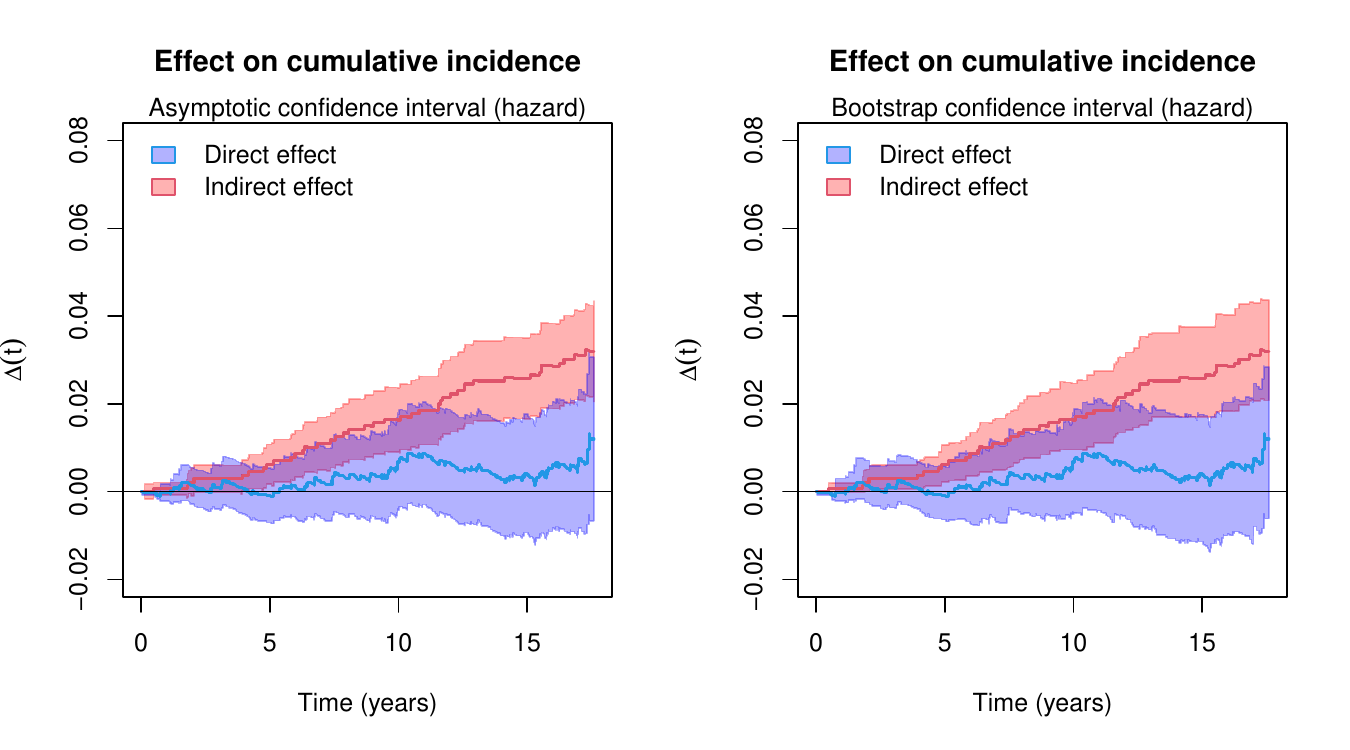}
\caption{Estimated treatment effects with 95\% confidence intervals for the hepatitis B data. Upper: Decomposition 1 (controlling the prevalence); lower: Decomposition 2 (controlling the hazard).} \label{fig_reveal_ci}
\end{figure}

Figure \ref{fig_reveal_ci} displays the 95\% confidence intervals of the treatment effects for the hepatitis B data, obtained by asymptotic formulas and bootstrap (200 resamplings) respectively. We find that the confidence intervals obtained by asymptotic formulas are wider than those obtained by bootstrap, but the substantial conclusions are consistent across these two methods.

\subsection{Transplantation type on mortality mediated by leukemia relapse}

There are two types of transplant modalities in allogeneic stem cell transplantation to cure acute lymphoblastic leukemia, namely haploidentical stem cell transplantation (haplo-SCT) from family and human leukocyte antigen matched sibling donor transplantation (MSDT). MSDT has long been considered as the first choice of transplantation because of lower transplant-related mortality \citep{kanakry2016modern}. However, recent findings indicated that haplo-SCT leads to lower relapse rate and hence lower relapse-related mortality especially in patients with positive pre-treatment minimum residual disease \citep{chang2020haploidentical}. We aim to study the causal mechanism of transplantation type on overall mortality mediated by leukemia relapse. The data include 303 patients undergoing allogeneic stem cell transplantation with positive pre-treatment minimum residual disease from an observational study \citep{ma2021an}. The treatment is transplantation type (1 for haplo-SCT and 0 for MSDT), the non-terminal event is relapse, and the terminal event is death. For illustration, we do not consider confounders at this moment.

\begin{figure}
\centering
\includegraphics[width=0.95\textwidth]{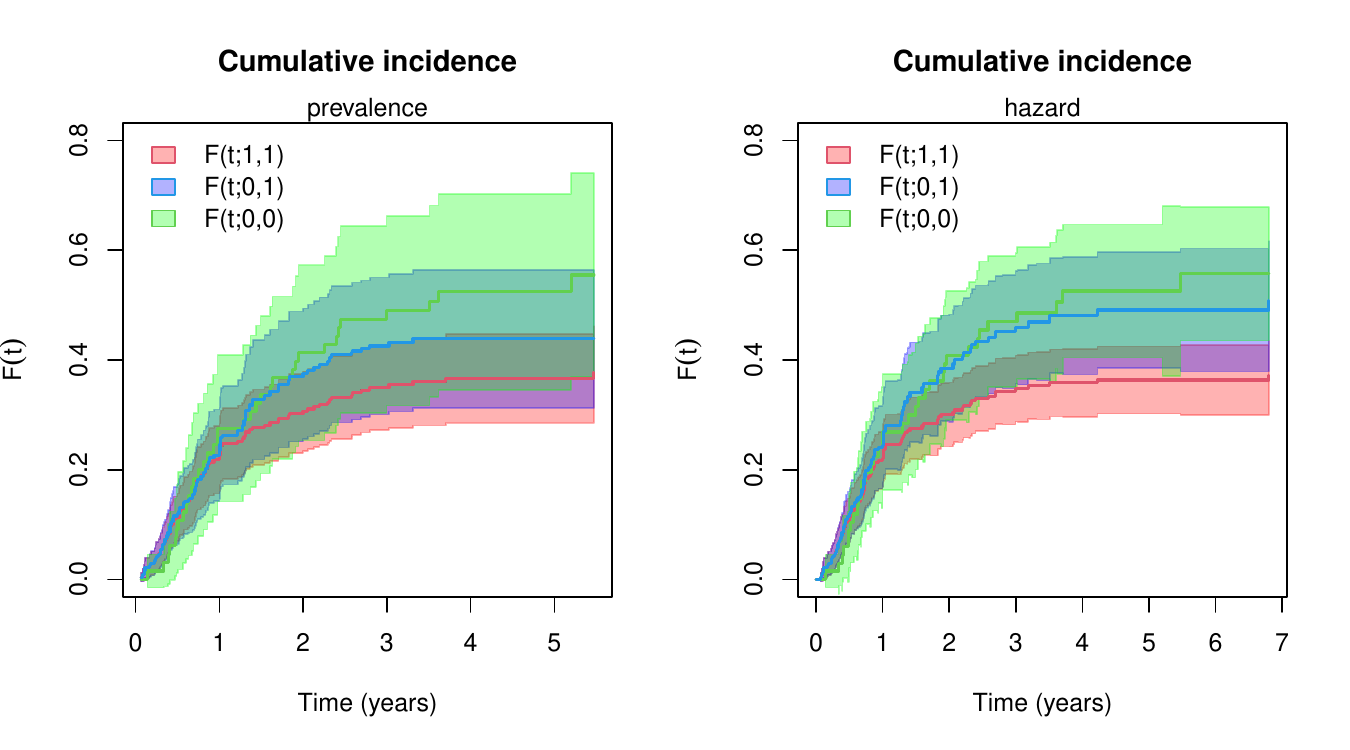} \\
\includegraphics[width=0.95\textwidth]{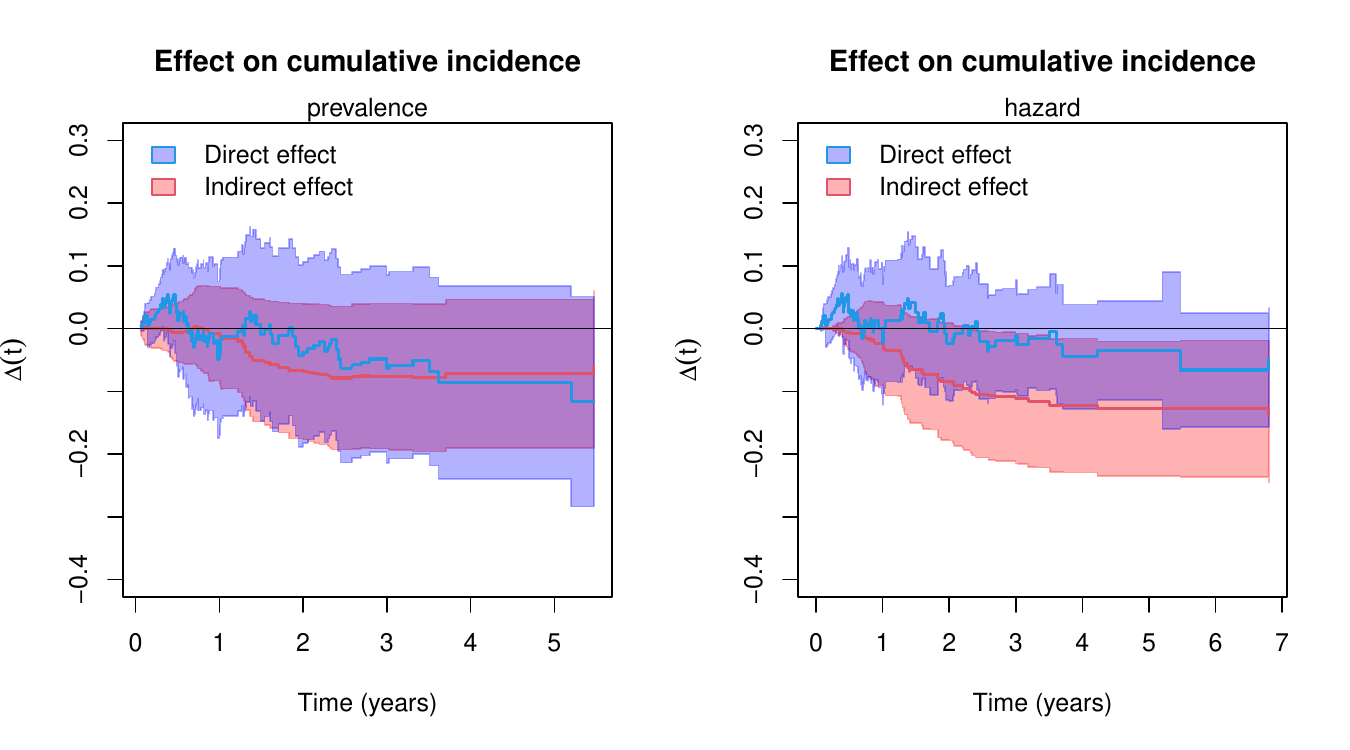}
\caption{Estimated counterfactual cumulative incidences and treatment effects with 95\% asymptotic confidence intervals for the leukemia data. Left: Decomposition 1 (controlling the prevalence); right: Decomposition 2 (controlling the hazard).} \label{fig_leukemia}
\end{figure}

Figure \ref{fig_leukemia} shows the counterfactual cumulative incidences of mortality and decomposed treatment effects with 95\% asymptotic confidence intervals (CIs). The left two figures decompose the total effect according to Decomposition 1 (controlling the prevalence), and the right two figures decompose the total effect according to Decomposition 2 (controlling the hazard). Interestingly, we observe different patterns for decomposed treatment effects under these two decompositions. Under Decomposition 1 (controlling the prevalence), the natural direct effect and natural indirect effect are both insignificant. Due to the limited sample size, the confidence intervals obtained by \citet{huang2021causal}'s method are wide, probably because the estimated variances are biased for a small sample size. However, under Decomposition 2 (controlling the hazard), the natural indirect effect is larger than the natural direct effect and becomes significant when time is large. We may conclude that haplo-SCT lowers the risk of mortality mainly through lowering the risk of relapse.

\begin{figure}
\centering
\includegraphics[width=0.95\textwidth]{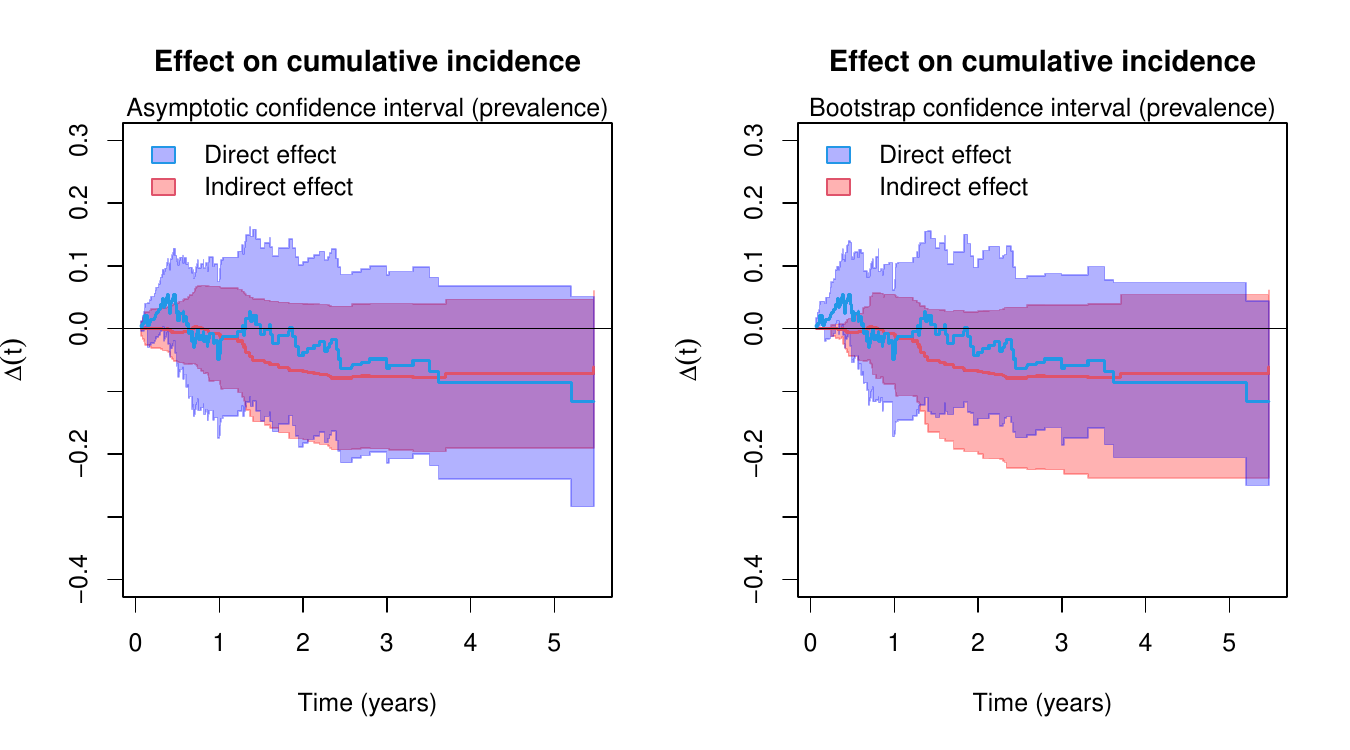} \\
\includegraphics[width=0.95\textwidth]{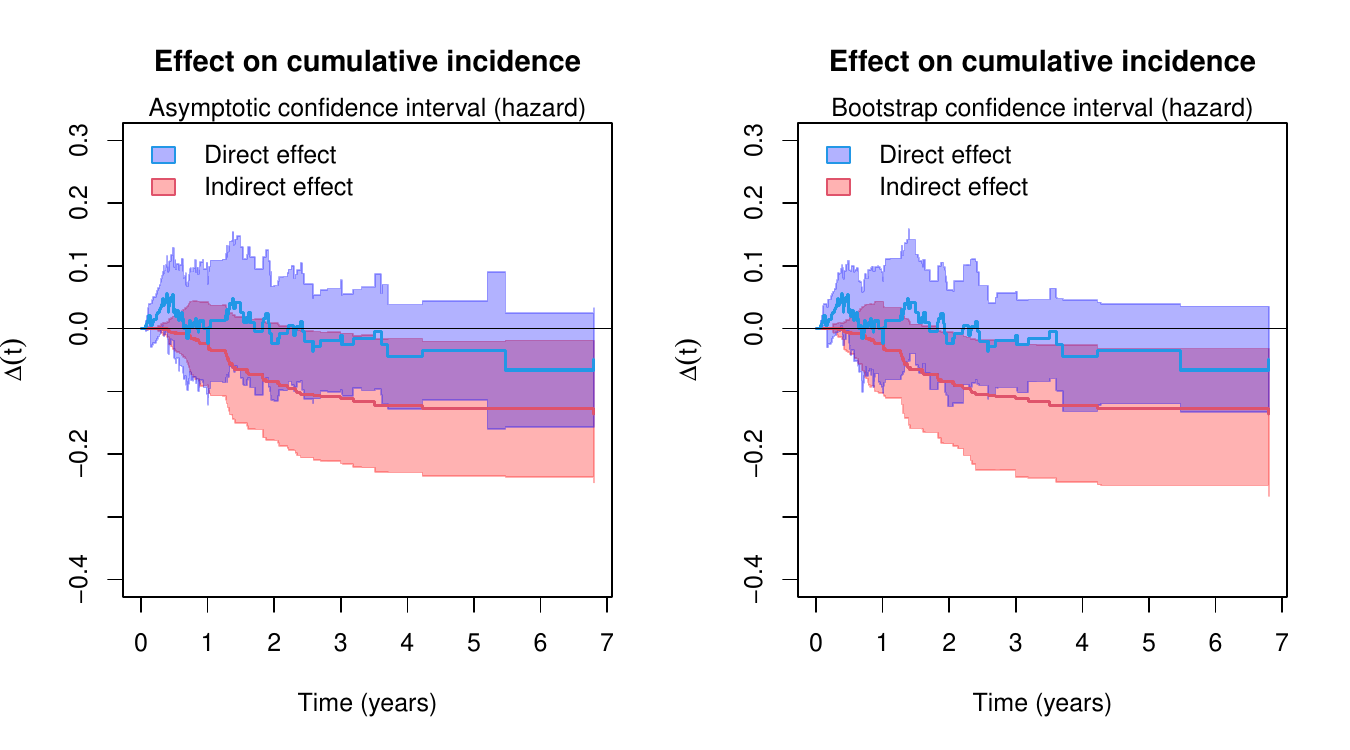}
\caption{Estimated treatment effects with 95\% confidence intervals for the leukemia data. Upper: Decomposition 1 (controlling the prevalence); lower: Decomposition 2 (controlling the hazard).} \label{fig_leukemia_ci}
\end{figure}

Figure \ref{fig_leukemia_ci} displays the 95\% confidence intervals of the treatment effects for the leukemia data, obtained by asymptotic formulas and bootstrap (200 resamplings) respectively. We have similar findings with those on the hepatitis data. Conclusions are the same across confidence interval methods, but are different across decomposition strategies.

The result by Decomposition 2 is easy to interpret from the clinical perspective. Haplo-SCT shows higher transplant-related mortality but lower relapse rate compared with MSDT \citep{kanakry2016modern}. A possible underlying biological mechanism is that haplo-SCT leads to stronger graft-versus-host disease (GVHD) and hence higher mortality since the human leukocyte antigen loci between the donor and receiver are mismatched. However, such kind of GVHD also kills residual leukemia cells and hence reduces the risk of relapse. This advantage of haplo-SCT compared with MSDT is recognized as the graft-versus-leukemia (GVL) effect \citep{wang2011superior, yu2020haploidentical}. Although this study is possibly subject to confounding (for example, by disease status, diagnosis type and age), the results still show some insight on the potential benefits of haplo-SCT. An analysis that addresses the confounding is considered in \citet{deng2023separable}, and the same conclusion is obtained.

Why does Decomposition 1 give insignificant result? When envisioning $F(t;0,1)$, Decomposition 1 tries to leave the hazard of death as natural while holding the prevalence of relapse among alive patients unchanged. This task is impossible due to the following reasons. Firstly, when switching the treatment from 0 (MSDT) to 1 (haplo-SCT), more individuals would experience transplant-related mortality, so the prevalence of relapse tends to get higher. Therefore, Decomposition 1 controlled the prevalence of relapse at a level lower than natural when calculating $F(t;0,1)$. Since relapse is strongly associated with relapse-related mortality, underestimation of the prevalence of relapse is reflected an underestimation of relapse-related mortality. Thus, the total incidence of mortality $F(t;0,1)$ is underestimated by Decomposition 1. This can be easily seen from the comparison of $F(t;0,1)$ in the first row of Figure \ref{fig_leukemia}.


\subsection{Natural direct and indirect effects, exchanging the treatment and control}

In the main text, we define the natural direct effect as $\Delta_{\DE}(t) = F(t;0,1) - F(t;0,0)$ and the natural indirect effect as $\Delta_{\IE}(t) = F(t;1,1) - F(t;0,1)$. Some literature argues that this kind of two-way decomposition neglects an interaction effect \citep{vanderweele2013three, vanderweele2014unification}. To avoid stepping into three-way or four-way decompositions, now we consider another two-way decomposition to assess our results:
\[
\Delta_{\DE}'(t) = F(t;1,1) - F(t;1,0), \ \Delta_{\IE}'(t) = F(t;1,0) - F(t;0,0).
\]
To estimate these effects, we just need to exchange the treatment and control, and calculate the negative effects.

\begin{figure}
\centering
\includegraphics[width=0.95\textwidth]{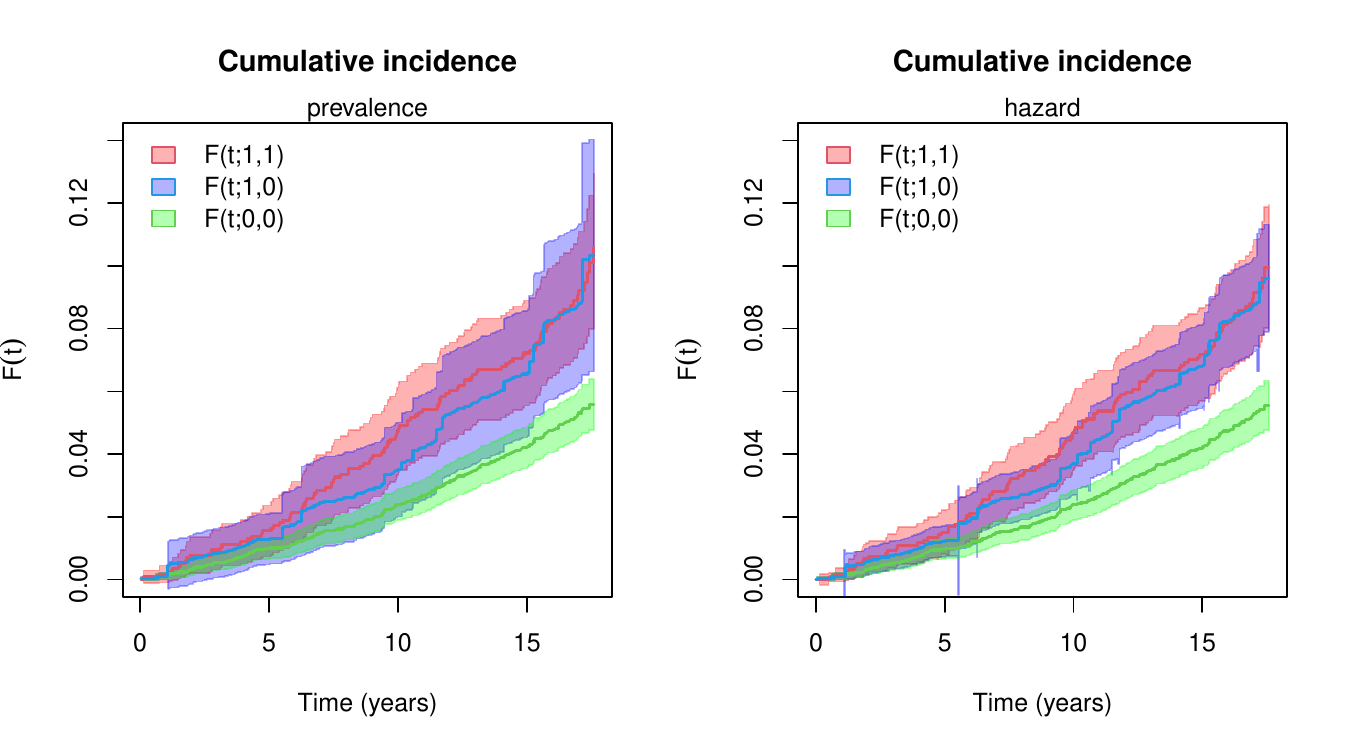} \\
\includegraphics[width=0.95\textwidth]{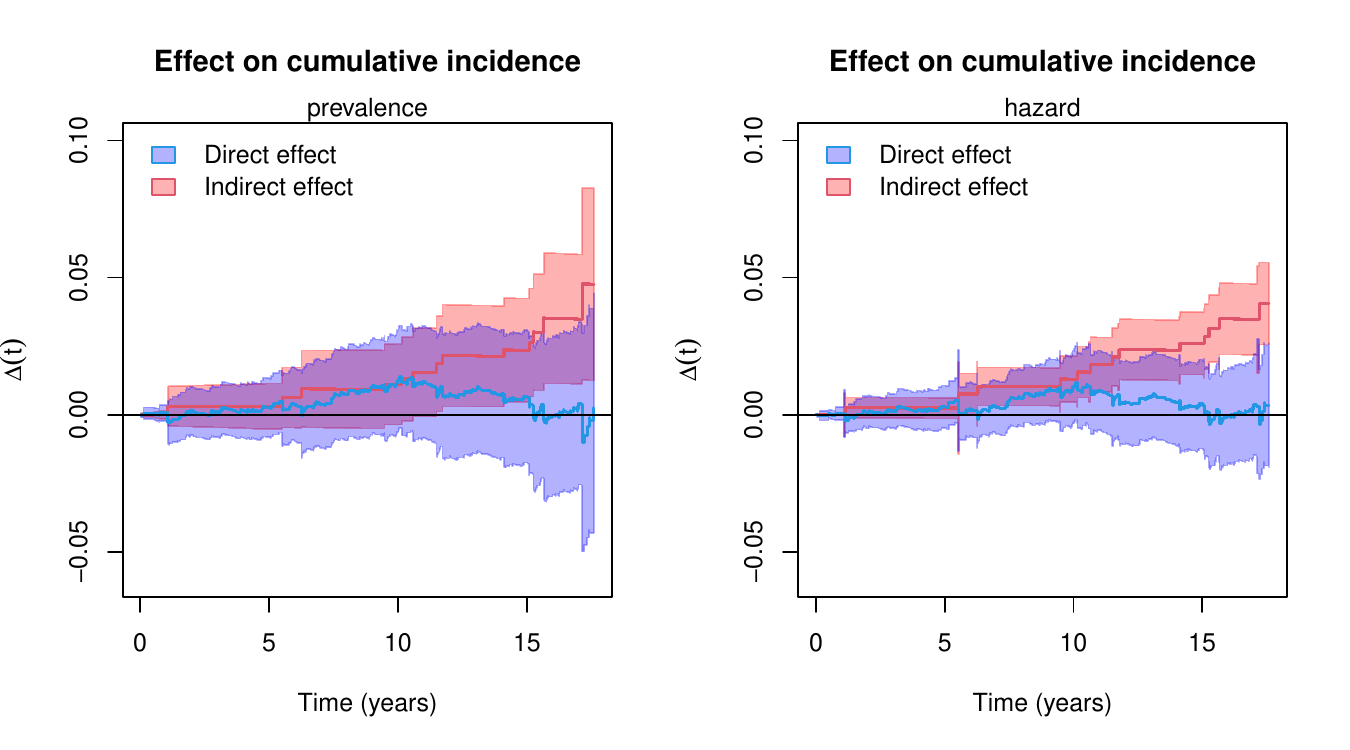}
\caption{Estimated counterfactual cumulative incidences and treatment effects with 95\% asymptotic confidence intervals for the hepatitis B data under the alternative two-way decomposition. Left: Decomposition 1 (controlling the prevalence); right: Decomposition 2 (controlling the hazard).} \label{fig_reveal_rev}
\end{figure}

Figure \ref{fig_reveal_rev} shows the counterfactual cumulative incidences of mortality and decomposed treatment effects with 95\% asymptotic confidence intervals for the hepatitis B data. The left two figures decompose the total effect according to Decomposition 1 (controlling the prevalence), and the right two figures decompose the total effect according to Decomposition 2 (controlling the hazard). The results on NDE and NIE are similar with those in the preceding sections.

\begin{figure}
\centering
\includegraphics[width=0.95\textwidth]{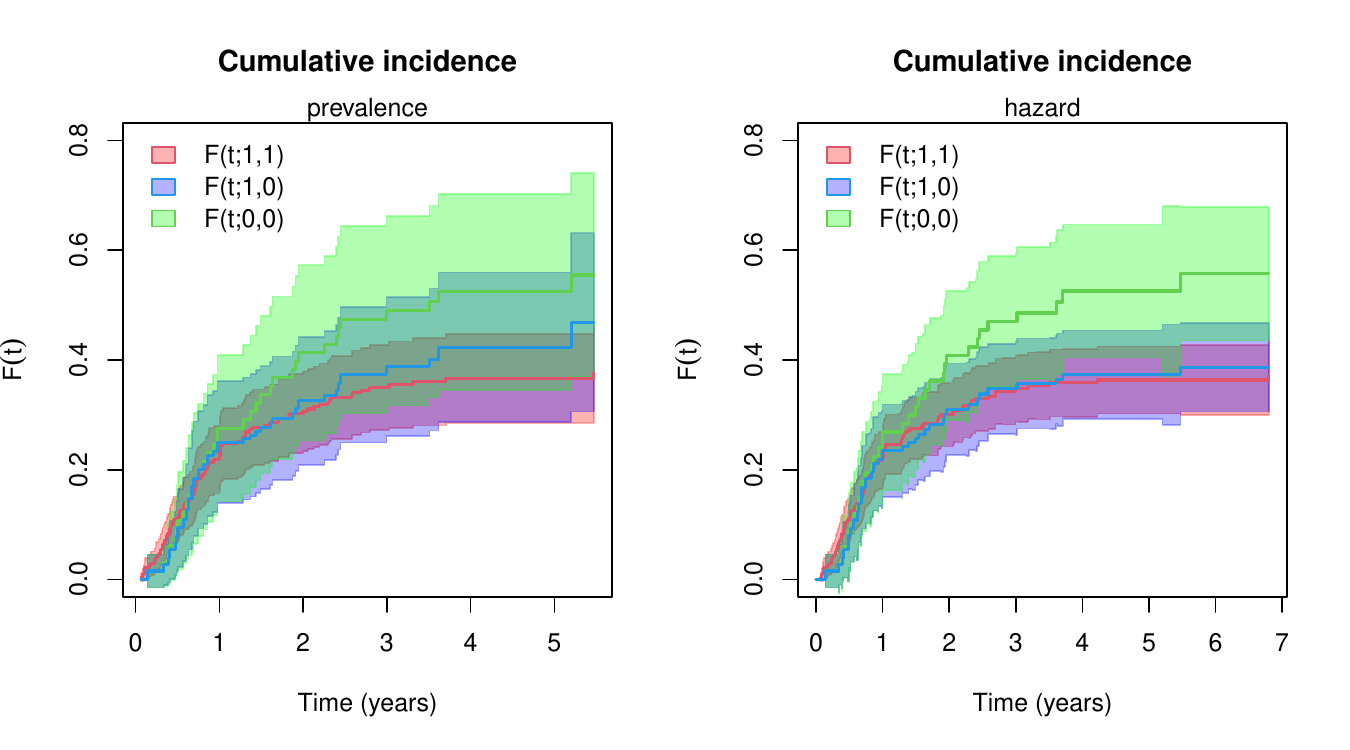} \\
\includegraphics[width=0.95\textwidth]{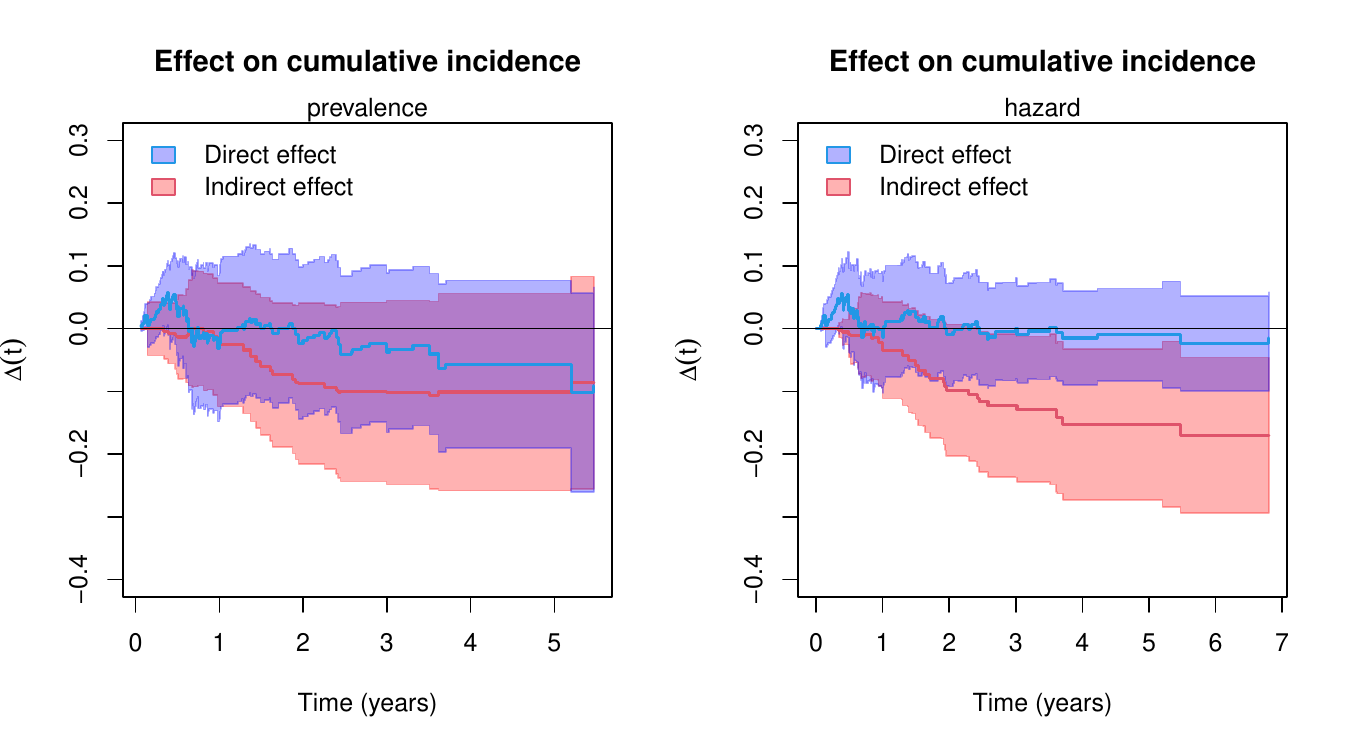}
\caption{Estimated counterfactual cumulative incidences and treatment effects with 95\% asymptotic confidence intervals for the leukemia data under the alternative two-way decomposition. Left: Decomposition 1 (controlling the prevalence); right: Decomposition 2 (controlling the hazard).} \label{fig_leukemia_rev}
\end{figure}

Figure \ref{fig_leukemia_rev} shows the counterfactual cumulative incidences of mortality and decomposed treatment effects with 95\% asymptotic confidence intervals for the leukemia data. The left two figures decompose the total effect according to Decomposition 1 (controlling the prevalence), and the right two figures decompose the total effect according to Decomposition 2 (controlling the hazard). The results on NDE and NIE are similar with those in the preceding sections.

In summary, the results remain consistent no matter how to decompose NDE and NIE. The substantial results from the real-data applications are still valid. Under the separable effects framework, \citet{deng2023separable} decompose the total effect into three separable pathway effects. Still using the leukemia data (and taking covariates into consideration), they find that the pathway effect from transplantation to relapse is significant, while the pathway effect from transplantation to transplant-related mortality and that from relapse to relapse-related mortality are insignificant.

\end{document}